\documentclass[amsmath,amssymb,aps,prd,floatfix,12pt,preprintnumbers]{revtex4}

\usepackage{bm}

\usepackage{amsfonts,amsbsy,latexsym,amssymb,amscd,amstext}
\usepackage{epsfig}
\usepackage{graphicx}
\usepackage{float}
\newcommand{\be}{\begin{equation}}
\newcommand{\ee}{\end{equation}}
\newcommand{\ba}{\begin{array}}
\newcommand{\ea}{\end{array}}
\newcommand{\baa}{\begin{array}}
\newcommand{\eaa}{\end{array}}
\newcommand{\bea}{\begin{eqnarray}}
\newcommand{\eea}{\end{eqnarray}}

\newcommand{\VO}{\hat{V}}
\newcommand{\VZ}{v_0}
\newcommand{\INT}{\prod_i\left(\int_0^1 dx^i\right)}
\newcommand{\modp}{||\vec{p}||_0}
\newcommand{\modk}{||\vec{k}||_0}
\newcommand{\phC}{\overline{\phi}}
\newcommand{\EC}{\mathcal{E}_C}
\newcommand{\ER}{\mathcal{E}_R}

\begin{document}

\title{Quantum corrections to vortex masses and energies}

\author{ Yago Ferreir\'os$^{a}$  and Antonio Gonz\'alez-Arroyo $^{b,c}$  \\
  $^a$ Instituto de Ciencia de Materiales de Madrid, CSIC \\
  Sor Juana In\'es de la Cruz 3, \\ Cantoblanco, 28049 Madrid, Spain.\\
  $^b$ Instituto de F\'{\i}sica Te\'orica UAM/CSIC,  \\
   Nicol\'as Cabrera 13-15, UAM\\
  Cantoblanco, 28049 Madrid, Spain.\\
  $^c$ Departamento de F\'{\i}sica Te\'orica, C-15 \\
       Universidad Aut\'onoma de Madrid, E-28049--Madrid, Spain \\
}
\email{yago.ferreiros@csic.es,  antonio.gonzalez-arroyo@uam.es}

\begin{abstract}
We study the 2+1 dimensional abelian Higgs model defined  on a spatial
torus at critical self-coupling. We propose a method to compute the
quantum contribution to the mass of the ANO vortex and to multi-vortex
energies. The one-loop quantum correction to  multi-vortex energies is 
computed analytically  at the critical value of the torus area (Bradlow limit).
For other values of the area one can set up an expansion around this
critical area  (Bradlow parameter expansion). The method is explained 
and the next-to-leading term explicitly evaluated. To this order, the 
resulting  energies  depend on the torus periods, but not on
the vortex positions. 
\end{abstract}



\preprint{IFT-UAM/CSIC-14-023;\  FTUAM-14-11}

\date{\today}


\maketitle



{\vskip 1cm}

\section{Introduction}
\label{s.intro}
Abrikosov-Nielsen-Olesen (ANO) vortices are string-like objects which appear as classical solutions
in  spontaneously broken  abelian gauge theories. They are important 
structures occurring in ordinary superconductors~\cite{abrikosov}, and 
corresponding solutions of the relativistic Abelian Higgs
model~\cite{Nielsen-Olesen}. Their stability has a topological origin
and this has triggered many theoretical and mathematical
works~\cite{taubes1}-\cite{taubes2}-\cite{jaffe}. It is simpler to view 
these vortices as solutions  in 2+1 dimensions.  The total magnetic flux
through space  is quantized 
and can be referred to as {\em vortex number}. The dimensionality of the 
space of solutions is twice the vortex number and can be interpreted as 
given by the two-dimensional positions  of a set of minimum-flux
vortices. A  particularly attractive  situation occurs at a critical value
of the Higgs self-coupling, at which   the gauge and Higgs field masses coincide.
The minimum energy equations reduce then to the first order Bogomolny
equations~\cite{Bogomolny}. 
At the classical level the vortex mass is given simply by the minimum
energy of the system with one unit of flux. It turns out that the
minimum  energy grows linearly with the vortex-number. This can be described by
saying that the interaction energy of vortices vanishes irrespective
of their relative positions. This is quite remarkable given the
non-linear character of the field equations. 

Although, many properties of the vortex solutions are known exactly,
there is no analytic expression for these solutions.  The single
vortex case is easy to describe numerically since the solutions are
rotational invariant~\cite{devega}. Thus, it can be expressed in terms of functions
of a single variable (the distance to its center). Much more difficult
is to obtain multi-vortex solutions numerically with a priori given
vortex centers\cite{weinberg}-\cite{rebbi}. In a previous paper~\cite{AGARAMOSI} one of the
present authors and Alberto Ramos derived a method to obtain analytic 
control of the solutions. This follows by considering the abelian Higgs
model on a 2-torus and expanding around a particular value of the area,
for which the solution is  known analytically. Considering enough terms
in the expansion one can obtain good approximations to the solution for
large torus sizes and even extrapolate to infinite size, where the 
solution tends to that of
the plane. What is more interesting is that the expansion can be
developed for multi-vortex solutions as well, and  for any location of the
vortex centers. Having analytic control allows many possible applications 
involving vortices. One such case  is to study vortex scattering, 
within the geodesic approximation~\cite{manton}, which
was done in Ref.~\cite{AGARAMOSII}. This leads to computations which,
even for the plane, are at least as precise as those obtained by other
numerical techniques.

In this paper we exemplify this idea even further by studying quantum
corrections to vortex masses and interaction energies. Given the
extension of the present work we will just explain the methodology and
compute the leading and next-to-leading terms in the expansion (known
as Bradlow parameter expansion). This fails too short for a reasonable
extrapolation to vortices on the plane. Extension to higher orders can
be done along the  same guidelines, with a straightforward but  technically
demanding effort. 

The  quantum energies of topological objects to one-loop order receive 
two types of contributions. First,  one has the Casimir energies, 
which follow by computing the difference of ground state energies 
between the topological non-trivial and trivial sector. This
subtraction should get rid of the most ultraviolet divergent
contributions, since topology is a global constraint. In addition, 
there are corrections to the classical energy due to the
renormalization of the lagrangian parameters. Both contributions are of 
order $\hbar$. 

Although the final result should be finite, at intermediate steps one 
will be manipulating divergent quantities. In this work, we have 
made use of the zeta-function regularization technique. This method 
is commonly used  in the literature of quantum corrections to
topological defects~\cite{Bordag}-\cite{Vassilevich}-\cite{Izquierdo}-\cite{Izquierdo2}-\cite{Izquierdo3}-\cite{Baacke}.
In particular, it is interesting to mention the study of
supersymmetric vortices in Refs.~\cite{Vassilevich} and~\cite{Rebhan}.
The situation here is much better than for ANO vortices  
because of the analytical control on the solutions, but 
also because Supersymmetry ensures cancellation of the contributions
of the bosons and fermions to the vacuum energy. Nevertheless, 
one still has to deal with the contributions coming from finite 
renormalization. 

For the purely bosonic case the calculation of the quantum mass of
self-dual vortices on the plane was addressed in
Ref.~\cite{Izquierdo}-\cite{Izquierdo2}-\cite{Izquierdo3}, 
using a mixture of  numerical and analytical techniques. The problem 
becomes easier to handle for circular invariant multivortices,
including the single vortex case. Nevertheless, the situation for spatially 
separated vortices is important, as it answers the question of whether 
quantum effects produce an attraction or repulsion, absent 
at the classical level. From that respect our methodology is much more
powerful, since one can fix the positions of the multiple vortices in 
any way and the analytical techniques apply equally well for all
situations.

For the rest of this section, we will describe the lay-out of this paper.
In  section~\ref{s.method} we particularize the abelian Higgs model to 
the case of a spatial  2-torus with arbitrary constant metric tensor 
and any value of the vortex 
number $q$. The metric can be parameterised in terms of the total area 
of the torus $\mathcal{A}$ and a complex parameter $\tau$ with positive 
imaginary part. We perform several manipulations to simplify the study 
of the classical and quantum system. In particular, we recall  that when 
the area attains a critical value $\mathcal{A}_c$ the classical
solutions become extremely simple. Furthermore, one can obtain
analytical control on the classical solutions in an expansion on the 
parameter $\epsilon=1-\mathcal{A}_c/\mathcal{A}$. This is essentially 
the Bradlow parameter expansion proposed in
Refs.~\cite{AGARAMOSI}-\cite{AGARAMOSII}. Our presentation here is slightly 
different, and in our opinion more elegant, than the one used in those
papers. 

For the study of the quantum system we use quantization in the $A_0=0$
gauge. This is the simplest and most appropriate for computing energies. 
In this gauge the physical Hilbert space is restricted to states that 
satisfy the Gauss constraint. Equivalently, physical states are those 
which are gauge invariant under the remaining time-independent gauge
transformations. Gauge invariance implies that, when studying the
spectrum of the quadratic fluctuations in the potential, gauge degrees
of freedom are zero-modes. Thus, they have a vanishing contribution to
the ground state quantum energy.  Thus, in computing the vacuum energy 
in a given topology, it is not necessary to fix the spatial  gauge and
no ghosts have to be added. 

The previous comments  become clear in the derivation of the
quantum energies for  critical area,  performed in section~\ref{s.critical}. The calculation 
is fairly simple, but we take advantage to present certain technical 
aspects necessary for the calculation at any order. In particular, the 
ideas explained earlier about the separation of gauge and non-gauge
degrees of freedom are easily checked. Finally, the Casimir energy
calculation employing the zeta-function regularization shows the
cancellation of the leading singularity, as expected. Indeed, this turns 
out to be the only singularity in the analytical continuation of the
energies. 

In the following section we explain the way in which the previous
result can be extended to other values of the area using the Bradlow 
parameter expansion. As an example we perform all the steps to produce 
the next-to-leading order correction to the masses. 
Part of the result depends on the calculation of 
the spectrum of the quantum fluctuation operator, which uses standard 
perturbative methods of Quantum Mechanics. The calculation  of 
the eigenvalues itself is presented in Appendix~\ref{calculation}. These
results are then used in combination with the zeta-function technique
to produce the Casimir energies to this order. 

The contribution of the quantum correction induced by the
renormalization of the parameters is performed in
Appendix~\ref{s.renormalization}. The results depend on the renormalization
prescription. A  prescription is adopted in which
the renormalization of the parameters is based on the behaviour of the
theory in the trivial topological sector and for large areas. This
makes sense, since typically one should not change the bare lagrangian of
the model when changing the area, the flat metric  or the vortex number. 
Thus, we set up a renormalization prescription within our $A_0=0$
context based on the behaviour of the effective potential under
space-time independent background fields. With the renormalization 
of parameters done in this way, we compute the counterterm contribution 
to the quantum energy which depends on the  area and on the number 
of vortices, but not on the location of these vortices or the metric
shape parameter $\tau$. With this result, all dependence of the quantum 
energies on these parameters should come from the Casimir energies
themselves. This dependence is finite and emerges from our calculation. 
This and other aspects  are analyzed in the concluding
section~\ref{s.conclusions}.

\section{The Abelian Higgs model on the torus}
\label{s.method}
In this section we will present the basic details of the model that we
are studying, and derive some of the formulas to be used later. 
We are considering the abelian Higgs model living on a two-dimensional 
spatial torus with non-vanishing flux. The lagrangian density of the model 
is given by 
\be
{\cal
L}=-\frac{1}{4}F_{\mu\nu}F^{\mu\nu}+\frac{1}{2}(D_\mu\phi)^*(D^\mu\phi)-\frac{e^2\lambda}{8}(|\phi|^2-v^2)^2
\ee
where $\phi(x)$ is a complex scalar field, henceforth referred as
Higgs field,  and $A_\mu(x)$ is the electromagnetic vector potential. 
In our notation the covariant derivative is
given by $D_\mu= \partial_\mu-i e A_\mu(x)$.

Usually the model is considered in flat Minkowski space, but it is
easily generalizable to compact spatial manifolds~\cite{bradlow}.
Here we will specialize to a two dimensional torus with flat metric. 
The torus is characterized by the two periods, given by two linearly independent
vectors. It is possible to perform   a linear change of coordinates and 
map the torus to a unit square with opposite sides identified. The 
points can then be parameterised by two real coordinates $0 \le x^i < 1$.
With this transformation the spatial euclidean metric tensor is mapped
onto a  constant metric tensor  $g_{i j}$. In these coordinates the 
area of the torus is $\mathcal{A}= \det(g) \equiv |g|$. As we will see,
the properties of the model become simple  at a particular value of
this area. One can study other values of the area by a systematic 
expansion method introduced in
Refs.~\cite{AGARAMOSI}-\cite{AGARAMOSII}, and referred as 
{\em  Bradlow parameter expansion}. In the aforementioned references
the expansion was carried to sufficiently high order so as to 
provide a good description of  multi-vortex classical solutions on an 
infinite plane. Having analytical control enables many other possible 
calculations.  Here we will show how the  formalism allows also to 
compute quantum corrections to the masses  analytically.
Our presentation differs slightly from the one used in
Ref.~\cite{AGARAMOSI}. The main difference is precisely the use 
of the square coordinates $x^i$. The Bradlow parameter
expansion can then be viewed as an expansion around a particular value 
$g^{(0)}$ of the constant Riemannian metric. This provides a more elegant 
formulation of the expansion.

Now let us now  introduce one complex vector $w^i$ such that
\be 
(w^i)^* w^j = g^{i j} + i \epsilon_{i j} I 
\ee
where we use the standard notation such that $g^{i j}$ are the
components of the inverse of $g$.
The $w^i$ vector is defined up to an overall phase.  However, one can
fix this freedom by imposing  $\Im(w^2)=0$, $\Re(w^2))>0$ (the symbols
$\Im$ and $\Re$ stand for the imaginary and real parts of a complex
number). With this choice one easily finds that  $I=\pm \frac{1}{\sqrt{|g|}}$ 
where $|g|$ is the determinant  of the metric tensor. The optimal choice of
sign is connected with the sign of the flux of the magnetic field
through the torus. From now on, without loss of generality,  we will
take this flux to be positive and correspondingly $I= \frac{1}{\sqrt{|g|}}$.

The Higgs field is to be seen as a section of a U(1) associated bundle
on the torus. The electromagnetic field is a connection on this bundle. 
As customarily done in the Physics literature, we will work with a 
 trivialization of the bundle. The Higgs field $\phi(x)$ can then be seen 
 as an ordinary complex function of the coordinates satisfying peculiar 
 boundary conditions:
 \be
 \phi(x+e_{(i)})=e^{i \vartheta_i(x)} \phi(x)
 \label{boundary}
 \ee
 where $e_{(i)}$ stands for the unit vector in the $i$th direction.
The topology of the bundle is encoded in the transition functions
$e^{i \vartheta_i(x)}$.

Now we construct the complex operator $D$ as follows
\be
D= w^1 D_1+ w^2 D_2 
\ee
where $w^i$ are the components of the complex vector introduced before. Let us compute 
\be
D^\dagger D= - (w^i)^* w^j D_i D_j = -  g^{i j} D_i D_j -
\epsilon_{i j} F_{ i j} \frac{e}{2\sqrt{|g|}} =- D^i D_i -
\frac{e B}{\sqrt{|g|}}
\ee
where the magnetic field is $F_{i j}= \epsilon_{i j} B$. Similarly we
arrive at 
\be
[D, D^\dagger]= \frac{2 e B}{\sqrt{|g|}}
\ee

One can use the previous definitions and results to re-express  the potential energy 
of the model with this metric. It is given by 
\be
\INT\  \left[ \frac{\sqrt{|g|}}{2} |D\phi|^2 +
\frac{e B}{2} |\phi|^2 +\frac{1}{2 \sqrt{|g|}}
B^2 + \frac{\sqrt{|g|}e^2\lambda}{8} (v^2-|\phi|^2)^2 \right]
\ee
The integrand can be  rewritten as 
\be
\label{integrand}
 \frac{\sqrt{|g|}}{2} |D\phi|^2  + \frac{1}{2 \sqrt{|g|}}( B +
\frac{e \sqrt{ |g|}}{2} (|\phi|^2-v^2))^2+ \frac{e B v^2}{2} +
\frac{\sqrt{|g|}e^2(\lambda-1)}{8}
(v^2-|\phi|^2)^2 
\ee
The integral of the third term is proportional to the flux of the 
magnetic field through the torus. The boundary conditions impose
that this flux is quantized $e \int B= 2 \pi q$, where $q$ is an integer 
called first Chern number of the bundle (vortex number in the Physics 
literature). Its value is determined in the 
choice of the transition functions $e^{i \vartheta_i(x)}$. 
Making use of a parity transformation we can always bring the flux and
$q$ to take positive values.  A look at the remaining terms of 
Eq.~\ref{integrand} shows that, at the critical value
$\lambda=1$, the potential energy attains its minimum for fields 
satisfying the Bogomolny equations:
\bea
D\phi&=&0\\B&=&\frac{e \sqrt{|g|}}{2}  (v^2-|\phi|^2)
\eea
From now on we will restrict ourselves to this  critical case.

One way to encode 
the flux condition is to write $B=\frac{2\pi q}{e} +\delta B$, where 
the integral of $\delta B$ over space vanishes. We will parameterize
the metric as $g_{i j}=\kappa g^{(0)}_{i j}$ and fix the normalization 
of the reference metric $g^{(0)}$ to satisfy  $\sqrt{|g^{(0)}|}=
\frac{4 \pi q}{v^2 e^2}\equiv {\cal A}_c$. Notice that the complex vector  $w_0^i$
associated to $g^{(0)}$ is related to the previous one by 
$w^i=\frac{1}{\sqrt{\kappa}}w_0^i$. In a similar fashion we can use the 
new vector $w_0^i$ to define $D^{(0)}$, connected to the previous one by 
$D=\frac{1}{\sqrt{\kappa}}D^{(0)}$. The advantage of our construction
is that of  keeping the dependence on the conformal factor $\kappa$ explicit. 
With these new definitions the potential energy (at the critical
coupling $\lambda=1$) takes the form:
\be 
\pi q v^2 + \INT\   \left[
\frac{2 \pi q}{e^2 v^2} |D^{(0)}\phi|^2  + \frac{v^2 e^2}{8 \pi q
\kappa}( \delta B + \frac{2 \pi q (1-\kappa)}{e}+
\frac{2 \pi q \kappa}{v^2 e} |\phi|^2)^2 \right]
\ee
The choice of scale of the reference  metric $g^{(0)}$ is such that, for 
$\kappa=1$, this potential energy takes its minimal value (equal 
to $\pi q v^2$) for $\delta B=\phi=0$. 
In other words, for the critical area (${\cal A}= {\cal A}_c=\frac{4 \pi q}{v^2
e^2} $) the solution of  the Bogomolny equations is very simple: 
vanishing Higgs field and constant magnetic field. 

We continue to fix our field redefinitions by rewriting the vector 
potential as
\be
A_i=A_i^{(0)} +\delta A_i
\ee
where  $A_i^{(0)}$ is a specific  vector potential leading  to
a constant magnetic field $\frac{2 \pi q}{e}$. Rather than working
with $\delta A_i(x)$ directly, we will be working with the 
complex field $\delta A(x)\equiv w_0^1 \delta A_1(x)+  w_0^2 \delta A_2(x)$ and its complex conjugate
$\delta A^*(x)$. With this choice, the complex covariant derivative $D^{(0)}$
can be written as
\be
 D^{(0)} = w_0^i (\partial_i-i e A_i^{(0)}- i e  \delta A_i) \equiv \tilde{D}- ie \delta A
 \ee
where the operator $\tilde{D}$ satisfies
\be
\label{tildeD}
[\tilde{D}, \tilde{D}^\dagger ]= e^2 v^2
\ee
which is, up to a scale, the commutation relation  of creation and annihilation
operators. Finally, one can express $\delta B$ in terms of $\delta A$
as 
\be
\delta B= -\frac{4 \pi q}{v^2 e^2} \Im(\tilde{\partial}\delta A^*)
\ee
with $\tilde{\partial}=w_0^1 \partial_1+ w_0^2 \partial_2$.

Before re-expressing the potential energy in terms of the two complex
fields $\delta A(x)$ and $\phi$, let us write down the kinetic term of
the Hamiltonian. In the $A_0=0$ gauge we have:
\be
\INT\, \frac{2 \pi q}{ e^2 v^2}\, \left[
\kappa |\dot{\phi}|^2 + |\delta \dot{A}|^2 \right]
\ee
It is convenient to eliminate the explicit dependence on the initial  metric 
by rescaling the fields in an obvious way:
\be
\phi\rightarrow \frac{\phi}{|g|^{1/4}}
\ee
\be
\delta A\rightarrow \frac{\delta A}{|g^{(0)}|^{1/4}}
\ee
The kinetic term then takes the canonical form 
\be
\INT\, \frac{1}{2}\, \left[
 |\dot{\phi}|^2 + |\delta \dot{A}|^2 \right]
\ee
To simplify notation we have preserved the symbols $\phi(x)$ and $\delta A$
in referring to the re-scaled fields. 

The final expression  of the potential energy after our field
redefinitions and  massaging is:
\be
\label{potential}
\pi q v^2 + \INT\   \frac{1}{2 \kappa} \left[
 |D^{(0)}\phi|^2  + ( - \Im(\tilde{\partial}\delta A^*)  - v  \sqrt{\pi q} (\kappa-1)+
\frac{v e^2}{ 4 \sqrt{\pi q}}  |\phi|^2 )^2 \right]
\ee
with $D^{(0)}= \tilde{D}- i\frac{e^2 v}{2 \sqrt{\pi q}} \delta A$. 
It is interesting to spend a few lines in explaining the
dimensionality and dependencies of the previous expression. Obviously
$v^2$ has dimensions of energy and provides its natural unit. Neither
$q$, $\kappa$ or the coordinates $x^i$ have dimensions. By our choice of 
coordinates, the dimensions of length square are transferred to the
metric tensor. Hence, both $\tilde{D}$ and $\tilde{\partial}$ have
dimensions of inverse length, whose natural unit is $ev$. Thus, 
the fluctuation operator and its eigenvalues will be
measured in $e^2 v^2$ units. This implies that the quantum contribution 
to the energies will become proportional to $ev$. This can be easily
understood if we realize that $ev\hbar c$ has dimensions of energy. 
From now on, however, we will continue to work in natural units
$\hbar=c=1$. Finally, the re-scaled background fields $\phi$ and 
$\delta A$ will appear naturally proportional to $1/e$. 

For later purpose it is convenient to write down an explicit parameterization 
of the critical area metric:
\be
\label{crit_met}
g^{(0)}=\mathcal{A}_c\,  \bar{g}(\tau)
\ee
where  $\mathcal{A}_c=\frac{4 \pi q}{v^2 e^2}$ is the critical area and 
$\bar{g}(\tau)$ is a conformally equivalent 
metric of unit determinant. The metric $\bar{g}(\tau)$ is dimensionless and 
can be parameterised  in terms of complex number $\tau$ 
as follows:
\be
\label{unit_det}
\bar{g}(\tau)= \frac{1}{\Im(\tau)}\begin{pmatrix} 1 & \Re(\tau) 
\\ \Re(\tau)& |\tau|^2 \end{pmatrix}
\ee
The two periods that define the torus can be changed into an
equivalent set (generating the same lattice) by a change of coordinates  
belonging to SL(2,$\mathbf{Z}$). This generates a class of equivalent 
metric tensors $\bar{g}$, which amount to transforming $\tau$ by an element of 
the modular group. Our results are then expected to be modular
invariant. We recall that the modular group is generated by two
transformations $T:\tau \longrightarrow \tau+1$ and  $S:\tau
\longrightarrow -1/\tau$. This allows us to restrict $\tau$ to a
fundamental domain, which can be chosen to be given by $|\tau|\ge  1$ 
and $|\Re(\tau)|\le \frac{1}{2}$.

Having presented the model and the basic fields, we will now briefly
describe how the vortex masses and energies are defined up to one loop
order. 
In each topological sector (characterized by a value of $q$),
we compute the minimum energy of the system. This has a classical
contribution corresponding to the potential energy of the
minimum energy solutions. In our case, the latter are the solutions of
the Bogomolny equations, and the corresponding energy is $\pi q v^2$. 
In addition, the ground-state energy receives quantum corrections. At one-loop
this correction follows by expanding  the hamiltonian  around classical 
solutions,  and keeping only  quadratic terms  in the fluctuations. 
The quadratic form  is given in terms of an  operator $\VO$, whose diagonalization 
defines the normal modes. Quantizing the system, we get a system of decoupled 
harmonic oscillators whose gap energies are given by the square root 
of the eigenvalues of the aforementioned operator. The ground-state
energy is the corresponding for the system of oscillators, namely
one half of the sum of the energies for 
each oscillator. As is well-known, this sum is divergent and some
regularization method is needed to handle the result.

The   vortex Casimir mass $\EC$ is defined as the difference of minimum
energies (classical + quantum) between the $q=1$ and $q=0$ sectors,
for the same metric. Multivortex Casimir energies  are equally defined,  by 
subtracting the energy of the zero-flux sector from that with  $q>1$. 
A priori this can depend not only on the metric parameter $\tau$ 
but also on the location of the vortices. However, it is expected that the
leading ultraviolet divergence is independent of these positions and 
on the value of $q$. Hence, the subtraction of the regularized 
quantities might be convergent,  or at least less divergent than the 
individual vacuum energies. This methodology is the standard one in 
computing finite Casimir energies. 

To compute the  vortex mass one must add an extra contribution, which
is also of order $\hbar$. This comes from quantum corrections to the 
classical energies, which are due to the renormalization of the
lagrangian parameters. We will refer to these extra contribution as 
counterterm mass  $\ER$.

The difficulty in carrying out the procedure described above to compute the 
vortex mass and multivortex energies is  that, in general, the solutions of 
the Bogomolny equations are not  known in closed analytic form. This forces the
program to be performed  numerically, as explained in the introduction. 
On the contrary, our method allows an analytic treatment based on  expanding 
the result around  particular  values of the metric associated to a  
critical value  for the area of the torus. For that value, as
mentioned earlier, the solutions of 
the Bogomolny equations are known and  very simple, allowing 
the whole program to be  carried over to completion using analytical 
techniques. This will be done in the next section. For other values of the
area of the torus, one can set up an expansion around the critical metric
and compute all the terms in the expansion in a systematic way. The
method is explained in  section~\ref{s.expansion}, and the leading 
order correction evaluated explicitly as an example.

\section{Vortex mass for the critical area case}
\label{s.critical}
To exemplify the whole procedure, let us consider here the case of critical 
area $\kappa=1$. As mentioned previously, the classical
solution for $q\ne0$  in this case is extremely simple: $\phi=\delta A= 0$. 
Thus,  the quadratic piece in the expansion of the potential is given by 
\be
\label{critpotential}
\prod_i (\int_0^1 dx_i)\   \frac{1}{2 } \left[
 |\tilde{D}\phi|^2  + (\Im(\tilde{\partial}\delta A^*))^2
 \right]\quad ,
 \ee
 where $\Im$ stands for imaginary part. 
 In the sector of Higgs fluctuations, the operator to diagonalize 
 is $\tilde{D}^\dagger \tilde{D}$. Comparison with the
 creation-annihilation operator algebra shows that the eigenvalues 
 of this operator are given by $e^2 v^2 n$ for all non-negative integers
 $n$. In the gauge potential sector the operator to diagonalize is 
 just obtainable in terms of $\tilde{\partial}$, which can
 diagonalized with plane waves $e^{i p_i x^i}$. The eigenvalues are given 
 by $|w_0^i p_i|^2=\modp^2=(g^{(0)})^{i j}p_i p_j$. 
 
Our previous discussion has been extremely naive. We have skipped
several relevant technicalities: boundary conditions, zero-modes and
gauge invariance. In the following subsections we will 
consider them in turn. In so doing we will develop the necessary
machinery to deal with the computation at any value of $\kappa$.

\subsection{Basis of Field space}
The question of boundary conditions is very relevant. The fields and
their fluctuations satisfy homogeneous boundary conditions.  
For the Higgs field $\phi(x)$, they are given in Eq.~\ref{boundary}. 
The space of fields satisfying these boundary conditions  defines a 
pre-Hilbert space  $\mathcal{H}_q$. We can follow  a standard quantum 
mechanical  formulation to study this space and the operators acting on it.
This formalism was developed extensively in
the appendix of Ref.~\cite{AGARAMOSI}, and here we will only review
the necessary results. The reader is addressed to that
reference for a detailed description.

One of the main results is that  the space $\mathcal{H}_q$ decomposes
naturally into $q$ orthogonal subspaces:
\begin{equation}
\mathcal{H}_q = \oplus_{s=0}^{q-1}\mathcal{H}_{q,s}
\end{equation}
The decomposition is associated to a symmetry group. This group
is a discrete subgroup of the translation group (combined with gauge
transformations).  The operators $\tilde{D}^\dagger$ and $\tilde{D}$ 
act on each of these subspaces without changing  the value of $s$. 
As mentioned previously, these operators satisfy the same algebra as
creation-annihilation operators up to a  multiplicative factor. Hence, 
we can introduce a basis of $\mathcal{H}_{q,s}$ using eigenstates of
the number operator $\tilde{D}^\dagger\tilde{D}/(e^2 v^2)$.
Incidentally, this basis is the same one that diagonalizes the fluctuation 
lagrangian for the critical area Eq.~\ref{critpotential}. The spectrum 
is given by $e^2 v^2 n$, for any non-negative integer $n$. Thus, the
result is the one anticipated previously, but now we know that each
eigenvalue is q-fold degenerate, corresponding to the different values
of $s$.  

In what follows, we will not need the explicit form of the basis 
states $\Psi_{n,s}(x)$, which satisfy the standard orthogonality
conditions 
\be
\INT\  \Psi^*_{n,s}(x)  \Psi_{n',s'}(x) =\delta_{n,n'}\delta_{s,s'}
\ee
Furthermore, the action of the operator $\tilde{D}$ on these states can be
read out trivially from the harmonic oscillator formulas
\bea
\tilde{D} \Psi_{n,s}(x)=e v \sqrt{n} \Psi_{n-1,s}(x)\\
\tilde{D}^\dagger \Psi_{n,s}(x)=e v \sqrt{n+1}  \Psi_{n+1,s}(x)
\eea
where $n\in \mathbf{Z}^+\cup\{0\}$ and $s=0,\ldots,q-1$.

Now we proceed to study the space of $\delta A$ fields. From the
previous considerations one concludes that they are periodic on the 
two-torus with period 1. To diagonalize the fluctuation hamiltonian 
at critical area, one can indeed choose a basis of plane waves 
$e^{i \vec{p}\vec{x}}$.  However, the boundary conditions impose that 
the momentum is given by  $\vec{p}=2 \pi \vec{k}$, where
$\vec{k}$ is a vector of integers. Plugging this plane-wave state into the
fluctuation formula Eq.~\ref{critpotential} we see that the
corresponding eigenvalue is $4 \pi^2 \modk^2$, 
where $\modk= \sqrt{g^{(0)\, i j} k_i k_j}$ as expected.

Thus, collecting the two results, we can write down the formula for the
quantum contribution to the  ground-state energy in the $q\ne0$ sector at critical area:
\be
q e v \sum_{n=0}^{\infty}\sqrt{n} +
\pi\sum_{k_1=-\infty}^\infty \sum_{k_2=-\infty}^\infty \modk
\label{zeroordermodes}
\ee
In getting to this formula we simply added one half of the square root of the
previously found eigenvalues of the fluctuation operator. There are some
subtleties, though, concerning the degeneracy of each eigenvalue.
The eigenvectors have to be taken to define a real vector space. Thus, 
the q-fold degeneracy of the Higgs fluctuation  potential turns into a 
2q-fold degeneracy for this vector space over the 
real numbers. For the gauge field part, each plane wave contributes a
single eigenvalue as we will explain below.

\subsection{Zero-modes and gauge invariance}
Zero-modes are eigenstates of eigenvalue zero of the fluctuation
operator. Although they do not contribute to the quantum mass, it is 
interesting to take a look at them to understand their origin. By
looking at Eq.~\ref{zeroordermodes} one sees, first of all,
that there are $2q$ zero-modes associated to $n=0$.  In addition,
there is another  zero mode  corresponding to constant vector potentials 
($\vec{k}=0$). These zero-modes reflect the dimensionality of the moduli
space of classical  solutions, which is $2q$ (see the analysis later on).

In addition, there are an infinite number of zero-modes associated to
gauge invariance. At this level, this  shows up 
in the fact that the potential depends on $\Im(\tilde{\partial}\delta
A^*)$. Half of the degrees of freedom drop out when taking the imaginary part. 
To separate gauge-dependent and gauge-invariant degrees of freedom 
it is necessary to modify the Fourier decomposition as follows:
\be
\delta A(x)=\sum_{k_1=-\infty}^\infty \sum_{k_2=-\infty}^\infty e^{i
2 \pi \vec{k}\vec{x}+ i  \alpha(\vec{k})}\, \hat{G}(\vec{k})
\label{four.decomp.}
\ee
where $\alpha(\vec{k})$ is defined by the expression 
\be
w_0^i k_i = e^{i \alpha(\vec{k})}  \modk
\ee
Now let us write $\hat{G}(\vec{k})=\hat{G}_1(\vec{k})+i \hat{G}_2(\vec{k})$
where $\hat{G}_i(\vec{k})=\hat{G}^*_i(-\vec{k})$. If we now 
substitute in the expression for  $\Im(\tilde{\partial}\delta
A^*)$, one sees that only $\hat{G}_1(\vec{k})$ appears in the
result. Thus, the  gauge degrees of freedom are associated to
$\hat{G}_2(\vec{k})$. This can also be seen by Fourier analyzing  a 
pure gauge term  $A_i(x)= \partial_i \varphi(x)$, and noticing that it has only
$\hat{G}_2(\vec{k})$ coefficients. 

Special treatment is required for the $\vec{k}=0$ modes of the vector
potential which, as commented earlier, also give rise to zero-modes.
Strictly speaking, these two  are  gauge-invariant modes. On the other
hand, one of the $n=0$ modes of the Higgs field is actually a gauge 
mode associated to global gauge transformations 
$\phi\longrightarrow e^{i \alpha} \phi$.
Thus, altogether we got  $2q+1$ gauge invariant zero-modes. This does
not match with the $2q$ parameters of the moduli. As we will see later, 
it turns out that one of the zero-modes is only accidentally so for 
critical area. The reader is addressed to Ref.~\cite{AGARAMOSI} for a 
more detailed explanation of the topology and dimensionality of the 
moduli space.

\subsection{Subtraction of the $q=0$ energy}
The calculation for vanishing flux is quite different. We show here
the result for an arbitrary value of the metric $g$. The minimum
energy solution is given by a constant Higgs field $\phi(x)=v$ and 
a vanishing vector potential $A_i(x)=0$, up to gauge transformations.
The quadratic fluctuation terms around this vacuum are well-known, being 
a simple example of the Higgs mechanism. In addition to gauge modes, 
the degrees of freedom correspond to 3 real massive fields.  One is
the real Higgs  field and the other two are the components of the
massive vector potential. At the critical value of the self-coupling 
($\lambda=1$), the mass of the photon and of the Higgs field are both
equal to $ev$. Hence, the vacuum energy becomes
\be
\label{vac_energy}
\frac{3}{2} \sum_{k_1=-\infty}^\infty \sum_{k_2=-\infty}^\infty
\sqrt{4 \pi^2||\vec{k}||^2 +e^2 v^2}
\ee
Notice that the formula is valid for every value of the constant metric
$g_{i j}$. The dependence appears through $||\vec{k}||^2=g^{i j} k_i
k_j$. 

As mentioned previously, the vortex Casimir energy $\EC$ is obtained  by
subtracting the $q=0$ vacuum energy from  the $q=1$ one. For larger
values of the flux ($q>1$) the same procedure leads to the  multivortex 
energy. Notice, however, that the same metric has to be used for the 
subtracted piece. Since the critical value of the area depends on $q$,
so will be the case for the vacuum energy  subtraction. In what follows
we will try to work as much as possible keeping the flux $q$ arbitrary,
and write down the final formulas to make this dependence explicit. 

In order to perform  a subtraction of two divergent  quantities we need to handle
them by some regularization procedure. 
Here we will use the method of analytical continuation, also known 
as zeta-function technique.  Let us explain the method in a generic 
way before applying it to our situation.

Let $\lambda_i$ denote the eigenvalues of the fluctuation operator
$\VO$ in increasing order. The quantum contribution to the ground-state energy
at one loop  is given by 
\be
{\cal E}_Q= \frac{ev}{2}\sum_{i=1}^\infty \sqrt{\lambda_i}
\ee
with $0<\lambda_i\le \lambda_j$ for $i<j$. We have factored out from
$\VO$ the quantity $ev$ having dimensions of mass and providing, as explained
earlier, the natural unit for quantum energies. The quantities $\lambda_i$ are hence
dimensionless.

Although, the previous  expression for ${\cal E}_Q$ is divergent,
we can define a function of the complex 
variable $s$  by\be
{\cal E}_Q(s) \equiv  \frac{ev}{2}\sum_{i=1}^\infty
(\lambda_i)^{\frac{1}{2}-s}
\ee
which will be convergent for $\Re(s)>s_0>0$. To make the expression
well-defined for $s>1/2$ one must, in addition,  exclude zero-modes from 
the sum.  Formally, the quantum energy is the analytical continuation of 
this function to $s=0$.  Obviously, the divergence of the initial one-loop 
energy reflects itself in the appearance of singularities as we move from 
the region of  analyticity to the point $s=0$. It could happen, however,
that if we subtract two divergent expressions, the corresponding $s$-dependent
functions are such that the divergences cancel each other, and one gets 
a smooth continuation. This necessarily happens whenever the initial
expression is finite. As we will see later, this is indeed the case 
for our vortex energies.

A good way to evaluate ${\cal E}_Q(s)$ is to rewrite it as 
\be
{\cal E}_Q(s)= \frac{ev }{2 \Gamma(s-1/2)}
\int_0^\infty dx\,  x^{s-3/2} \, \sum_i e^{-x\lambda_i}=
\frac{ev}{2
\Gamma(s-1/2)}
\int_0^\infty dx\,  x^{s-3/2} \, \mathrm{Tr}(e^{-x \VO}) 
\ee
where the  operator $e^{-x \VO}$ is called the heat-kernel of
the operator $\VO$. For $x>0$ its trace  is well defined. The 
divergence at $s=0$ appears because the trace  does not vanish strongly
enough as $x\longrightarrow 0$ to make the integration convergent at
the lower limit.

Let us apply these relations  to  the case of the multivortex energy at 
critical metric (or area). The one-loop  quantum energy at non-trivial 
topology ${\cal E}_Q(s)$ is given by Eq.~\ref{zeroordermodes}. It is
the sum of two terms. The analytical continuation of the first one,
coming from  the  Higgs field fluctuations,  can be easily
recognized as $qev\zeta(s-1/2)$, where $\zeta(x)$ is Riemann zeta-function.
This function is analytic for $\Re(s)>3/2$, is well-defined at $s=0$,
and has a simple pole with unit residue at $s=3/2$. The second term,
coming from the vector potential, can be defined using the 
corresponding heat kernel, whose trace is 
\be
\mathcal{F}(x/q)=\sum_{k_1=-\infty}^\infty \sum_{k_2=-\infty}^\infty e^{-x\xi}
\ee
where the dimensionless quantity $\xi$ is given by 
\be
\xi\equiv\frac{4 \pi^2}{e^2v^2} \modk^2 = \frac{4 \pi^2}{e^2v^2}
(g^{(0)})^{i j} k_i k_j =  \frac{\pi}{q}\bar{g}^{i j}(\tau) k_i k_j
\ee
and $\bar{g}(\tau)$ is the conformally equivalent metric of unit
determinant introduced earlier.   
Notice that the function
$\mathcal{F}(x)$ is indeed equal  to the 2-dimensional Riemann theta 
function $\Theta(z,\Omega)$, as given for example in Ref.~\cite{tata},
for $z=0$ and  $  \Omega= -i x  (\bar{g}(\tau))^{-1} $. The properties 
of this function  realize the invariance under transformations of the
modular group  in the complex parameter $\tau$.

Using the definition of $\mathcal{F}(x)$ and combining it with the  
$\zeta$-function term,   we get 
the analytical continuation of the one-loop quantum energy given by 
\be
{\cal E}^{(0)}_Q(s)= evq \zeta(s-1/2)+ \frac{ev q^{s-1/2}}{2 \Gamma(s-1/2)}
\int_0^\infty dx\,  x^{s-3/2}
\, (\mathcal{F}(x)-1)
\ee
where the superscript $(0)$ recalls that the result is valid for the
critical area case.  The function is well-defined for $\Re(s)>3/2$. 
The second term also
develops a pole at $s=3/2$. This can be deduced using the modular
invariance of the theta function, which implies that the leading
behaviour of $\mathcal{F}(x)$ for  $x\longrightarrow 0$ is
$$ \mathcal{F}(x) = (\det(\Omega))^{-1/2}+ \ldots = \frac{1}{x} +\ldots $$
where the dots represent terms with  powers of the exponential  of $-1/x$. To
display the  singularity explicitly, we can add and subtract a term 
$\frac{1}{x}e^{-x}$ to the integrand. One gets 
\be
{\cal E}^{(0)}_Q(s)= evq  \zeta(s-1/2)+ \frac{ev q^{s-1/2}}{2 (s-3/2)} +
\frac{ev q^{s-1/2}}{2 \Gamma(s-1/2)}
\int_0^\infty dx\,  x^{s-3/2}
\, (\mathcal{F}(x)-1-\frac{ e^{-x}}{x})
\ee
where the integral is well defined for $\Re(s)>-\frac{1}{2}$.

A similar treatment can be done for the one-loop vacuum energy in the trivial
topology sector. This time, however,  we will do the calculation for an 
arbitrary value of the metric $g=\kappa g^{(0)}$. Using
Eq.~\ref{vac_energy} and the previous definitions we get 
\be
\label{calc_vac}
{\cal E}_\varnothing(s)=\frac{3 e v}{2} \sum_{\vec{k}}
(\frac{\xi}{\kappa}+1)^{1/2-s}= \frac{3 e v (\kappa q)^{s-1/2}}{2
\Gamma(s-1/2)}
\int_0^\infty dx\,  x^{s-3/2} e^{-\kappa q x} \mathcal{F}(x)
\ee
To explicitly display the singularity of the integral  we might add and 
subtract $1/x$ from $\mathcal{F}(x)$ to get to 
\be
{\cal E}_\varnothing(s)=\frac{3 e v \kappa q}{2 (s-3/2)} +  \frac{3 e v (\kappa q)^{s-1/2}}{2
\Gamma(s-1/2)}
\int_0^\infty dx\,  x^{s-3/2} e^{-\kappa q x} (\mathcal{F}(x)
-\frac{1}{x})
\ee
The integral part is now an entire function  and the only singularity resides 
in the single pole at $s=3/2$. The fact that the residue is
proportional to $\kappa$ shows that there is a divergent contribution to the
energy which is extensive and, hence, proportional to the area. We
expect  a similar behaviour for the most divergent contribution to the
energy for non-trivial topology.  

Coming back to the critical area case, we can set $\kappa=1$ in the
previous formula to get ${\cal E}^{(0)}_\varnothing(s)$, which should 
be subtracted from ${\cal E}^{(0)}_Q(s)$. One sees that the pole at $s=3/2$
cancels out in the difference, and the whole expression becomes regular 
down  to $\Re(s)=-1/2$. 
According to our previous considerations, we interpret this as evidence
that the vortex  and multivortex Casimir energies are indeed finite
quantities, and their value can be obtained by setting $s=0$ in the
difference. The  result is 
given by
\bea
\nonumber
\EC^{(0)}&=&ev(q\zeta(-1/2)+q-\frac{1}{3\sqrt{q}})\\
\label{crit_energy}
&+& \frac{ev}{2 \sqrt{q}\Gamma(-1/2)}
\int_0^\infty dx\,  x^{-3/2} \left(\mathcal{F}(x)(1-3e^{-qx})
-1+\frac{1}{x}(3e^{-qx}-e^{-x})\right)
\eea
As commented in the introduction, the quantum  vortex energy results
from   adding to this result the $\hbar$ contribution to the 
classical energy $\ER^{(0)}$. 
This follows from the renormalization of the parameters in the
lagrangian. This extra term, however, depends on the renormalization
prescription that is adopted. In appendix~\ref{s.renormalization} we 
set up a prescription that comes out quite natural within our formulation.
Combining Eq.~\ref{crit_energy} with this result (Eq.~\ref{ERFinal}) we get 
\be 
\label{totE0}
\mathcal{E}=\EC^{(0)}-\frac{39}{32}q ev 
\ee
The new term changes the numerical value but has no influence on the 
dependence of our result on  $q$ and $\tau$,  that we will 
now analyze.

The quantum contribution to the vortex mass can be obtained by 
setting $q=1$ in Eq.~\ref{totE0}.
The terms which do not involve an integral add up to  $-0.75997$. On
the other hand, the integral concentrates all the dependence on the
metric parameter $\tau$.  It attains its minimum value $0.169259$ 
for $\tau=e^{i \pi/3}$, which adds up to a quantum vortex mass of $-0.590711$
in $ev$ units. Going back  to the euclidean metric coordinates, 
we can see that this value of $\tau$ corresponds to the periods 
characteristic of a triangular lattice  of vortices. The value of the
energy corresponding to a square lattice ($\tau=i$) is $-0.589877$, 
which gives a very small difference. It is remarkable that the quantum
correction gives rise to  a minimum energy configuration which coincides
to the vortex lattice  obtained in type II superconductors. 

The situation changes considerably for other values of $q$. At large
values, the dominant contribution comes from the first two terms in
Eq.~\ref{crit_energy}. This produces a linear dependence with 
$q$ and slope  $-0.426636\ ev$. Adding up the classical energy, which is
also linear in $q$, we get $\mathcal{M}=\pi v^2 -0.426636\ ev$.
The latter value is,  henceforth, the energy  per vortex on a large 
vortex situation, and  can 
be interpreted as an alternative estimate of the vortex mass. Clearly
the quantum contribution becomes sizable when $e/v$ is large enough. 

Corrections to the linear  behaviour are proportional to $1/\sqrt{q}$, 
up to exponentially suppressed terms. The coefficient  of the $1/\sqrt{q}$ term is
dominated by the photon energy contribution and depends on 
the metric parameter $\tau$. This time, however, it is maximal  at 
$\tau=e^{i \pi/3}$ and minimal along the line $\tau=i r$, becoming
negative divergent at infinite $r$. Using properties of integrals of 
the Jacobi theta function one can calculate the  coefficient of the 
$1/\sqrt{q}$ term, up to terms exponentially suppressed in $r$ to be
\be
-\frac{r^{3/2}}{\sqrt{q}} \frac{\zeta(3)}{4 \pi^{3/2}}-
\frac{r^{-1/2}}{\sqrt{q}} \frac{\sqrt{\pi}}{12}
\ee
For $r\sim 6 q$ this term becomes comparable with the linear term in
$q$.

\section{Bradlow parameter expansion of quantum energies}
\label{s.expansion}
In the previous section we evaluated the quantum correction to the 
vortex mass on a spatial  torus of critical area. In this section we
will show how it is possible to extend this result to other values of 
the area. This is done by setting up a power expansion in the conformal
factor $(\kappa-1)$.  The methodology has been  used previously in
Ref.~\cite{AGARAMOSI} to obtain an analytic expression for the
multivortex field configurations at critical coupling: solutions of the
Bogomolny equations. With sufficient high order calculations one
obtains a competitive description of the solutions on the plane.
An advantage of these analytic expressions is that they facilitate other 
calculations involving vortices, such as their scattering
behaviour~\cite{AGARAMOSII}. Here we will explain how one can 
set up a similar expansion for the quantum corrections to the vortex 
masses. The leading term is given by the result of the previous section 
and the next to leading term will be computed in the present one. 

Before going into details, let us enumerate briefly the different steps 
involved in the procedure. As explained previously, the one-loop quantum 
contribution follows by calculating the spectrum of the operator
$\VO$ determining the quadratic fluctuations around classical solutions of
the equations of motion. In our case, these are just the solutions of
the Bogomolny equations. It is precisely in this step where one makes
use of the results of Ref.~\cite{AGARAMOSI}, by obtaining a  
series expansion of these solutions in powers of the square root of
$\epsilon=(\kappa-1)/\kappa$. Substituting these background fields onto the 
expression of the fluctuation operator, one arrives to  an equivalent
expansion for this operator
\be
\VO=\sum_{n=0}^\infty \epsilon^{n/2} \VO^{(n)}
\ee
Its eigenvalues  also admit  an expansion 
\be
\lambda_i = \lambda^{(0)}_i + \sum_{p=1}^\infty \epsilon^p
\lambda^{(p)}_i
\ee
involving only integer powers of $\epsilon$. The coefficients
$\lambda^{(p)}_i$ can be obtained  
applying  the standard technique, analogous  to that employed 
in Quantum Mechanics when using perturbation theory. 

The final step is to plug this result into the analytically continued 
formulas for the ground-state quantum energy $\mathcal{E}_Q(s)$ and expand the
result in powers of $\epsilon$ to give 
\be
{\cal E}_Q(s)= \frac{e v }{2 \Gamma(s-1/2)} \int_0^\infty dx\,  x^{s-3/2}
\,
\sum'_i e^{-x\lambda^{(0)}_i} (1+ \sum_{p=1}^\infty \epsilon^p
\sum_{l=1}^p \frac{(-x)^l}{l!} c_i(p,l))
\ee
where
\be
c_i(p,l) =\sum_{k_1,k_2,\ldots,k_l=1}^\infty \delta(\sum k_a=p)
\prod_{a=1}^l \lambda^{(k_a)}_i
\ee
Finally, as done before, the total quantum energy is obtained  
by subtracting the contribution of trivial topology, extrapolating
to $s=0$ and adding the counterterm contribution. 

In the following subsections we will apply the above procedure to 
the calculation of the quantum vortex energies  to order $\epsilon$.
In that case $p=l=1$ and the coefficient $c_i(1,1)=\lambda_i^{(1)}$. 
This allows us to circumvent an important complication arising when
there are degenerate levels at lowest order. If degeneracy is
accidental it is broken by higher order corrections. This implies 
a diagonalization procedure within the subspace associated to the same 
lowest order eigenvalue $\lambda_i^{(0)}$. However, for the calculation of the 
mass to order $\epsilon$, all we need is the sum of the  $\lambda_i^{(1)}$
within that space, i.e. the trace of the operator in the degenerate
space. This avoids the much more involved problem of computing the
splittings and eigenstates.

\subsection{Solutions of the Bogomolny equations}
In terms of our main complex fields $\phi$ and $\delta A$, the
Bogomolny equations can be read off from the form of the potential
Eq.~\ref{potential}:
\bea
\label{Bogo_eqone}
\tilde{D} \phi&=& i\frac{e^2 v}{2 \sqrt{\pi q}} \delta A \phi \label{bog11}\\
\label{Bogo_eqtwo}
\Im(\tilde{\partial}\delta A^*)&=&- v  \sqrt{\pi q} (\kappa-1)+
 \frac{v e^2}{ 4 \sqrt{\pi q}}  |\phi|^2\label{bog12}
\eea

For $\kappa=1$  the solution is given by $\phi=\delta A=0$. Hence, the
idea is  simple: express  the solution as a power series in
$\sqrt{\kappa - 1}$ or $\sqrt{\epsilon}=\sqrt{\kappa - 1}/\sqrt{\kappa}$.
The coefficients of this expansion can be
obtained iteratively.  The occurrence of the square root of 
$\kappa - 1$ can be understood if we integrate over space the
second Bogomolny equation. One gets 
\be
\int d^2x\, |\phi|^2=\VZ^2
 (\kappa-1)
 \ee
 where $\VZ=\frac{ \sqrt{ 4 \pi q}}{e}$. 
 From this equation one concludes that the Higgs field $\phi$ is
 proportional to $\VZ\sqrt{\kappa - 1}$. Apart from this normalization
 factor, all the remaining corrections involve integer powers of
 $\epsilon$. For example, notice that the left hand side of
 Eq.~\ref{Bogo_eqtwo} is of order $\epsilon$. Hence, $\delta A$ starts at
 order $\epsilon$. Plugging this in Eq.~\ref{Bogo_eqone} one gets a
 correction of order $\epsilon^{3/2}$ to $\phi$, and so on and so forth.
 In summary, $\delta A$ can be expanded in integer powers of $\epsilon$ 
 and $\phi$ in  half-integer powers. 
 
Obviously, in finding the solutions one must take proper care of the
boundary conditions. The appropriate boundary conditions are valid for 
the solutions as well as for the fluctuations, and were explained in
the previous section. The $\phi$ field belongs to the space
$\mathcal{H}_q$, and hence can be expanded in the basis $\Psi_{n,s}$.
On the other hand, $\delta A$ is periodic on the torus and can be
expanded in our modified Fourier expansion(Eq.~\ref{four.decomp.}) 
in terms of coefficients $\hat{G}_1(\vec{k})$ and $\hat{G}_2(\vec{k})$. 

One might wonder about how does the multiplicity of the Bogomolny 
solutions arises. First of all, one can fix the gauge by requiring that 
the coefficients $\hat{G}_2(\vec{k})$ vanish. Another freedom  is
associated with translation invariance and this can be fixed by
setting \\
 $\hat{G}_1(\vec{k}=0)=0$. The remaining multiplicity is fixed, as we 
 will see,  by fixing the lowest order terms in the expansion. This is
 one of powers of this method, which allows obtaining multivortex  
 solutions with arbitrary centers. 
 
In any case, it is not our purpose to describe in detail the methodology 
to get results to higher order in $\epsilon$, since that was the
subject of Ref.~\cite{AGARAMOSI}. In that reference we gave two
procedures to compute the corrections. One method, described in the appendix 
of that paper, uses the terminology of a quantum mechanical
description, despite the fact that the problem is indeed classical.
The pre-Hilbert space is given by the space of Higgs fields
$\mathcal{H}_q$, a basis of which is provided by the functions
$\Psi_{n,s}(x)$. One can define operators
$\hat{U}(\vec{k})$ acting on $\mathcal{H}_q$,  and amounting to
multiplication by $e^{2 \pi i \vec{k}\vec{x}}$. The matrix elements of
these operators on the basis states will be the main formula needed to
perform all the calculations.  The result, which we reproduce here, is
\bea
\nonumber
X_{m n}^{s' s}(\vec{k})\equiv \langle m,s'| \hat{U}(\vec{k}) |n,
s\rangle&=& \int d^2 x\  \Psi^*_{m,s'}(x) e^{2 \pi i \vec{k}\vec{x}}
\Psi_{n,s}(x)= \\
 U_{s' s}(\vec{k}) e^{i(m-n)\beta}(-1
)^{(M+n)}e^{-\frac{\xi}{2}}\xi^{\frac{|m-n|}{2}}
&\times&\sum_{j=0}^M(-1)^j\frac{\sqrt{n!m!}\xi^j}{j!(M-j)!(j+|m-n|)!}=
\label{matrixel}\\
\nonumber 
 U_{s' s}(\vec{k}) e^{i(m-n)\beta}(-1
 )^{(M+n)}e^{-\frac{\xi}{2}}\xi^{\frac{|m-n|}{2}}
 &\times& \frac{\sqrt{M!}}{\sqrt{(M+|m-n|)!}}\, L^{(|m-n|)}_M(\xi)
\eea
where  $M=min(m,n)$,  $\beta=\alpha(\vec{k})+\pi/2$ and $\xi\propto
\modk^2$ is the quantity introduced in the previous section. The
function $L^{(|m-n|)}_M(x)$ denotes a generalized Laguerre polynomial. All the 
$s,s'$ dependence sits in the unitary $q\times q$ matrix $U(\vec{k})$,
given by 
\be 
U_{s' s}(\vec{k})=\delta_{s's+k_1}e^{-i\pi k_1k_2/q}e^{-2\pi i s k_2/q}
\ee
where the $\delta$ is to be taken modulo $q$. 
For the single vortex case (q=1) this is just $e^{-i\pi k_1k_2}$.
The symbols $X_{m n}^{s' s}$ can be regarded as the components of a
matrix  $X$ which  satisfies  $X(-\vec{k})=X^\dagger(\vec{k})$.

Let us conclude this subsection by performing the program up to
leading order in $\epsilon$.  The
solution for $\phi(x)$ to lowest  order is given by the solution of
Eq.~\ref{bog11} with vanishing right hand-side. This is proportional to
the ground state of the corresponding harmonic oscillator:
\begin{equation}
\phC(x)= \sqrt{\epsilon} \VZ  \sum_{s=1}^q c_s \Psi_{0,s}(x)
\end{equation}
where the constants $c_s$ are the components of a q-dimensional
complex vector of unit norm. These constants encode the multiplicity of
the solutions and are related to the position of the vortices (see
Ref.~\cite{AGARAMOSI}).  For the single vortex case ($q=1$)  $c_1$ is just a
phase.

To solve for $\delta A$
to this order, we simply have to substitute the previous expression 
in the right hand-side of Eq.~\ref{bog12} and use Eq.~\ref{four.decomp.} 
and Eq.~\ref{matrixel} to obtain
\begin{equation}
\bar{G}_1(\vec{k})=-\frac{\epsilon \VZ}{2 \sqrt{\xi}}\, e^{-\frac{\xi}{2}} u(\vec{k})
\end{equation}
where $u(\vec{k})=c_t^* U_{t s}(-\vec{k}) c_s$ carries the dependence
on the multivortex moduli parameters.  Notice that the equation only 
constrains $\bar{G}_1$.  Taking $\bar{G}_2$ to zero
amounts to a choice of gauge for the background field solution, which we will adopt.
In addition, for simplicity we also fix to zero the component associated to
$\vec{k}=0$. Results do not depend on these choices.

\subsection{Spectrum of Quantum Fluctuations}
Having found the minima of the potential in the last subsection, we now 
expand the fields around these solutions as follows
\be
\phi \longrightarrow \phC+\varphi
\ee
\be
\delta A\longrightarrow \overline{\delta A}+\delta a
\ee
and plug these into the expression of the lagrangian  keeping only
terms  quadratic in the quantum fluctuation fields $\varphi$ and
$\delta a$. 
In our case the result is 
\be
\delta V= \prod_i(\int_0^1 dx^i)\, \frac{1}{2 \kappa}\left[
|(\tilde{D}-i C\overline{\delta A})\varphi-iC \phC\delta
a|^2 +(-\Im(\tilde{\partial}\delta a^*) +
\frac{C}{2}(\phC\varphi^*+\phC^*\varphi))^2\right]
\ee
where  we introduced  the  constant $C=ev/\VZ$. Indeed, the whole potential is
proportional to $\frac{e^2v^2}{\kappa}$ so we can write (in a rather symbolic notation) 
\be
\delta V=\frac{e^2v^2}{2\kappa} (\langle\varphi|,\langle \delta a|)
\VO  \begin{pmatrix}|\varphi\rangle \\ | \delta a\rangle
\end{pmatrix}
\ee
in terms of a hermitian operator $\VO$, acting on the space
of  fluctuation fields $\varphi$ and $\delta a$, associated respectively
to the Higgs field and the vector potential. 
Our goal is  to  obtain the eigenvalues $\lambda_i$ of this operator. 
To do so, we substitute the background fields  ($\phC(x)$ and
$\overline{\delta A}(x)$) by their expansion in powers of
$\sqrt{\epsilon}$ derived in the previous section. We then  obtain 
\be
\VO=\VO^{(0)}+\epsilon^{1/2}\VO^{(1/2)}+\epsilon\VO^{(1)}+\mathcal{O}(\epsilon^{3/2})
\ee
We might  use indices to specify on which of the
fluctuation fields is the operator acting:
\be
\VO = \begin{pmatrix}\VO_{11} & \VO_{12} \\
\VO_{21} & \VO_{22} \end{pmatrix}
\ee
Since $\phC$ and $\overline{\delta A}$ are expandable in
odd and even powers of $\sqrt{\epsilon}$ respectively, one concludes
that the off-diagonal terms  ($\hat{V}_{12}=\hat{V}_{21}$)   and
the diagonal ones ($\hat{V}_{ii}$)  have the same property respectively.
From this, one easily concludes that the eigenvalues are expandable 
in integer powers of $\epsilon$:
\be
 \lambda=  \sum_{m=0}^\infty \epsilon^m \lambda^{(m)}
 \ee
The coefficients $\lambda^{(m)}$ can be determined by standard quantum
mechanical techniques to be described below.

\begin{figure}[H]
        \centering
\setlength{\unitlength}{0.240900pt}
\ifx\plotpoint\undefined\newsavebox{\plotpoint}\fi
\sbox{\plotpoint}{\rule[-0.200pt]{0.400pt}{0.400pt}}%
\begin{picture}(1500,900)(0,0)
\sbox{\plotpoint}{\rule[-0.200pt]{0.400pt}{0.400pt}}%
\put(150.0,131.0){\rule[-0.200pt]{4.818pt}{0.400pt}}
\put(130,131){\makebox(0,0)[r]{-0.15}}
\put(1419.0,131.0){\rule[-0.200pt]{4.818pt}{0.400pt}}
\put(150.0,252.0){\rule[-0.200pt]{4.818pt}{0.400pt}}
\put(130,252){\makebox(0,0)[r]{-0.1}}
\put(1419.0,252.0){\rule[-0.200pt]{4.818pt}{0.400pt}}
\put(150.0,374.0){\rule[-0.200pt]{4.818pt}{0.400pt}}
\put(130,374){\makebox(0,0)[r]{-0.05}}
\put(1419.0,374.0){\rule[-0.200pt]{4.818pt}{0.400pt}}
\put(150.0,495.0){\rule[-0.200pt]{4.818pt}{0.400pt}}
\put(130,495){\makebox(0,0)[r]{ 0}}
\put(1419.0,495.0){\rule[-0.200pt]{4.818pt}{0.400pt}}
\put(150.0,616.0){\rule[-0.200pt]{4.818pt}{0.400pt}}
\put(130,616){\makebox(0,0)[r]{ 0.05}}
\put(1419.0,616.0){\rule[-0.200pt]{4.818pt}{0.400pt}}
\put(150.0,738.0){\rule[-0.200pt]{4.818pt}{0.400pt}}
\put(130,738){\makebox(0,0)[r]{ 0.1}}
\put(1419.0,738.0){\rule[-0.200pt]{4.818pt}{0.400pt}}
\put(150.0,859.0){\rule[-0.200pt]{4.818pt}{0.400pt}}
\put(130,859){\makebox(0,0)[r]{ 0.15}}
\put(1419.0,859.0){\rule[-0.200pt]{4.818pt}{0.400pt}}
\put(150.0,131.0){\rule[-0.200pt]{0.400pt}{4.818pt}}
\put(150,90){\makebox(0,0){ 0}}
\put(150.0,839.0){\rule[-0.200pt]{0.400pt}{4.818pt}}
\put(365.0,131.0){\rule[-0.200pt]{0.400pt}{4.818pt}}
\put(365,90){\makebox(0,0){ 10}}
\put(365.0,839.0){\rule[-0.200pt]{0.400pt}{4.818pt}}
\put(580.0,131.0){\rule[-0.200pt]{0.400pt}{4.818pt}}
\put(580,90){\makebox(0,0){ 20}}
\put(580.0,839.0){\rule[-0.200pt]{0.400pt}{4.818pt}}
\put(795.0,131.0){\rule[-0.200pt]{0.400pt}{4.818pt}}
\put(795,90){\makebox(0,0){ 30}}
\put(795.0,839.0){\rule[-0.200pt]{0.400pt}{4.818pt}}
\put(1009.0,131.0){\rule[-0.200pt]{0.400pt}{4.818pt}}
\put(1009,90){\makebox(0,0){ 40}}
\put(1009.0,839.0){\rule[-0.200pt]{0.400pt}{4.818pt}}
\put(1224.0,131.0){\rule[-0.200pt]{0.400pt}{4.818pt}}
\put(1224,90){\makebox(0,0){ 50}}
\put(1224.0,839.0){\rule[-0.200pt]{0.400pt}{4.818pt}}
\put(1439.0,131.0){\rule[-0.200pt]{0.400pt}{4.818pt}}
\put(1439,90){\makebox(0,0){ 60}}
\put(1439.0,839.0){\rule[-0.200pt]{0.400pt}{4.818pt}}
\put(150.0,131.0){\rule[-0.200pt]{0.400pt}{175.375pt}}
\put(150.0,131.0){\rule[-0.200pt]{310.520pt}{0.400pt}}
\put(1439.0,131.0){\rule[-0.200pt]{0.400pt}{175.375pt}}
\put(150.0,859.0){\rule[-0.200pt]{310.520pt}{0.400pt}}
\put(794,29){\makebox(0,0){n}}
\put(171,495){\makebox(0,0){$+$}}
\put(193,186){\makebox(0,0){$+$}}
\put(214,817){\makebox(0,0){$+$}}
\put(236,771){\makebox(0,0){$+$}}
\put(257,517){\makebox(0,0){$+$}}
\put(279,313){\makebox(0,0){$+$}}
\put(300,238){\makebox(0,0){$+$}}
\put(322,280){\makebox(0,0){$+$}}
\put(343,397){\makebox(0,0){$+$}}
\put(365,543){\makebox(0,0){$+$}}
\put(386,670){\makebox(0,0){$+$}}
\put(408,743){\makebox(0,0){$+$}}
\put(429,745){\makebox(0,0){$+$}}
\put(451,680){\makebox(0,0){$+$}}
\put(472,569){\makebox(0,0){$+$}}
\put(494,443){\makebox(0,0){$+$}}
\put(515,332){\makebox(0,0){$+$}}
\put(537,259){\makebox(0,0){$+$}}
\put(558,236){\makebox(0,0){$+$}}
\put(580,263){\makebox(0,0){$+$}}
\put(601,330){\makebox(0,0){$+$}}
\put(623,422){\makebox(0,0){$+$}}
\put(644,520){\makebox(0,0){$+$}}
\put(666,609){\makebox(0,0){$+$}}
\put(687,676){\makebox(0,0){$+$}}
\put(709,714){\makebox(0,0){$+$}}
\put(730,721){\makebox(0,0){$+$}}
\put(752,699){\makebox(0,0){$+$}}
\put(773,654){\makebox(0,0){$+$}}
\put(795,595){\makebox(0,0){$+$}}
\put(816,528){\makebox(0,0){$+$}}
\put(837,463){\makebox(0,0){$+$}}
\put(859,405){\makebox(0,0){$+$}}
\put(880,360){\makebox(0,0){$+$}}
\put(902,329){\makebox(0,0){$+$}}
\put(923,316){\makebox(0,0){$+$}}
\put(945,317){\makebox(0,0){$+$}}
\put(966,333){\makebox(0,0){$+$}}
\put(988,359){\makebox(0,0){$+$}}
\put(1009,394){\makebox(0,0){$+$}}
\put(1031,434){\makebox(0,0){$+$}}
\put(1052,477){\makebox(0,0){$+$}}
\put(1074,519){\makebox(0,0){$+$}}
\put(1095,559){\makebox(0,0){$+$}}
\put(1117,594){\makebox(0,0){$+$}}
\put(1138,623){\makebox(0,0){$+$}}
\put(1160,644){\makebox(0,0){$+$}}
\put(1181,656){\makebox(0,0){$+$}}
\put(1203,659){\makebox(0,0){$+$}}
\put(1224,652){\makebox(0,0){$+$}}
\put(1246,636){\makebox(0,0){$+$}}
\put(1267,611){\makebox(0,0){$+$}}
\put(1289,579){\makebox(0,0){$+$}}
\put(1310,542){\makebox(0,0){$+$}}
\put(1332,501){\makebox(0,0){$+$}}
\put(150.0,131.0){\rule[-0.200pt]{0.400pt}{175.375pt}}
\put(150.0,131.0){\rule[-0.200pt]{310.520pt}{0.400pt}}
\put(1439.0,131.0){\rule[-0.200pt]{0.400pt}{175.375pt}}
\put(150.0,859.0){\rule[-0.200pt]{310.520pt}{0.400pt}}
\end{picture}
\caption{We display $(\delta\lambda_n-2)/(2 \sqrt{n})$ as a function of
$n$ for $q=1$ }
			        \label{fig.1}
				\end{figure}
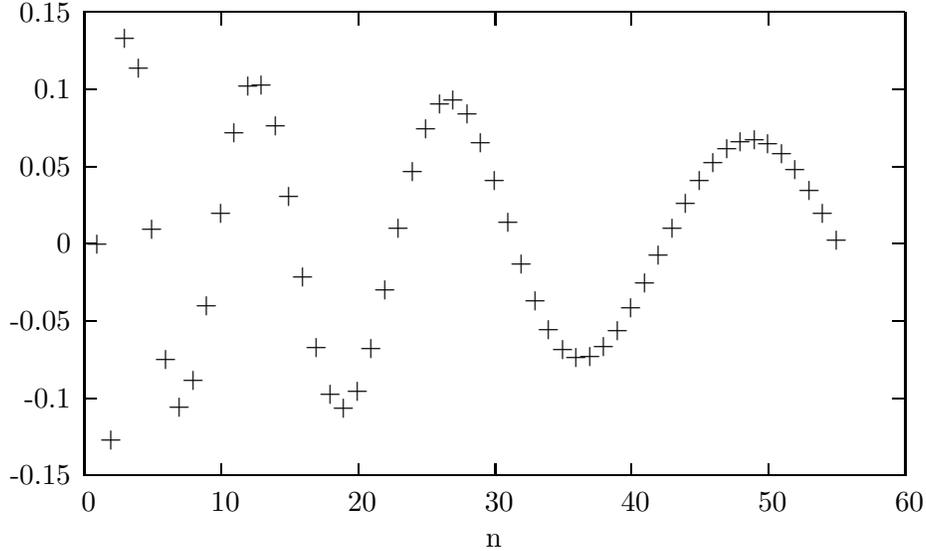

The first step is to diagonalize $\VO^{(0)}$, the operator
corresponding to critical area. This was done in the previous section.
The eigenstates of $\hat{V}_{11}^{(0)}$ will be labeled
$|n,s,\sigma\rangle$ and correspond to the functions $\Psi_{n, s}(x)$
for $\sigma=+$ and $i\Psi_{n, s}(x)$ for $\sigma=-$. As mentioned
earlier, we have to consider two states ($\sigma=\pm 1$) because we want to 
work with real vector spaces.  The corresponding eigenvalue is $n$,
and does not depend on $s$ or $\sigma$. Thus, to lowest order its  degeneracy 
is 2q, but this  might be broken by higher order corrections. Nevertheless,
for the purpose of computing the next  correction all we need is the 
trace of $\VO$ in the space characterized by eigenvalue $n$: $\delta
\lambda_n$. Applying standard perturbative techniques the linear correction
in $\epsilon$ is given by  
\be
\label{lambdan}
\delta \lambda_n = \epsilon  \sum_{s=0}^{q-1} \sum_{\sigma=\pm 1} \langle
n,s, \sigma | ( \VO_{11}^{(1)}+
\VO_{1 2 }^{(1/2)} (n-\VO^{(0)}_{2 2})^{-1} \VO_{21}^{(1/2)} ) | n, s, \sigma \rangle
 \ee
To facilitate the reading of the paper we will collect the calculation in
Appendix~\ref{calculation}, and give here  only the final result:
\be
\label{higgs_eigen}
 \delta \lambda_{n\neq0}= 
 \epsilon\left( 2qn-\sum_{j=0}^{n-2} \rho(j) \right)
 \ee
 where 
  \be
 \rho(n)=\sum_{\vec{k}}
 e^{-\xi} \frac{\xi^{n}}{n!}
 \ee
It is easy to see that $\rho(j)$ oscillates around a constant value of
$q$, with oscillations that are damped with increasing $q$. Hence,
$\delta \lambda_n$ should oscillate around $2q$. For $q=1$ the first
few values are displayed in Fig.\ref{fig.1}, showing the oscillatory
pattern.

A similar procedure applies for the eigenvalues associated to the vector 
potential. The  eigenfunctions of $\hat{V}_{22}^{(0)}$ will be labelled 
$|\vec{k}, L,\sigma \rangle$ and $|\vec{k}, T,\sigma\rangle$, for
positive $\vec{k}$. The label $\vec{k}$ corresponds to the modified Fourier
modes Eq.~\ref{four.decomp.}, with $T$ and $L$ referring to the gauge
invariant and gauge dependent parts, associated  to the coefficients
$G_1(\vec{k})$ and $G_2(\vec{k})$ respectively. To work with real
coefficients we take only half of the Fourier modes and split 
the coefficients $G_i(\vec{k})$ into its real and imaginary parts 
labelled by the symbols $\sigma=\pm 1$ respectively. Thus, the vectors 
$\vec{k}$ are restricted to positive values, meaning $k_1>0$ or
\{$k_1=0$ and $k_2>0$\}. Summing  up, the corresponding
eigenfunctions are given by 
\be
\label{chidef}
\chi^\eta_{\vec{k},\sigma}(x)\equiv P_\eta \frac{1}{\sqrt{2}}e^{i\pi (1-\sigma)/4} \left(e^{i 2 \pi \vec{k}\vec{x}+ i \alpha(\vec{k})} +\sigma e^{-i 2 \pi
\vec{k}\vec{x}+ i \alpha(-\vec{k})}\right) 
\ee
where $\eta=T,L$ for transverse or longitudinal photons
respectively, while $P_T=1$ and $P_L=i$. To lowest order, the transverse
photons $|\vec{k}, T,  \sigma \rangle$
have eigenvalue $\xi$,  while the longitudinal photons have eigenvalue 0. 

Calculating the correction to order $\epsilon$ follows the same steps
as for the Higgs. Once more we benefit from having to compute only the
trace within each degenerate sector. Thus the goal is 
\be
\label{lambdak}
\delta \lambda_{\vec{k}}  = \epsilon  \sum_{\sigma= \pm 1} \langle
\vec{k}, T, \sigma | (\VO_{22}^{(1)}+
\VO_{2 1 }^{(1/2)} (\xi-\VO^{(0)}_{1 1})^{-1} \VO_{12}^{(1/2)} ) | \vec{k}, T, \sigma\rangle
 \ee
The details of the calculation are collected in 
Appendix~\ref{calculation}. The result is very simple
\be
\delta \lambda_{\vec{k}\ne 0} = 2 \epsilon
\ee

One can also compute the correction to the zero eigenvalue. Not
surprisingly the correction  vanishes, since it is associated to 
a gauge symmetry which is valid at all orders in $\epsilon$. Anyhow, 
it serves as a check of our manipulations. 
 
We have left out from the previous spectrum corrections to  the values at 
$n=0$ and  $\vec{k}=0$.  Together with the gauge modes of the
longitudinal photons ($G_2(\vec{k})$), these were  zero-modes at 
leading order.  We emphasized then that $2q$  of these zero-modes 
are associated to the moduli of solutions of the Bogomolny equation. Since
this holds for any value of the area, they should remain zero modes at
any order in our $\epsilon$ expansion.  Thus,  as mentioned earlier, 
2 of the $2q+2$ zero-modes cannot be gauge-invariant zero-modes. 
Indeed, we already mentioned that one of them is associated with 
global gauge transformations. The remaining zero
mode was just accidentally so at critical area and is broken at order
$\epsilon$. Indeed, in the appendix we found $\delta \lambda_{n=0}=1$
and we explained that it is  associated to a
multiplicative re-scaling of the solution, $c_s\longrightarrow t c_s$
with $t$ real. Since this is a single state, we will leave it out from
the analytical continuation and add its contribution to the final result.

\subsection{Vortex energies to order $\epsilon$}
\label{s.vortex_energy}

In this subsection we will use the results of the previous subsection 
to compute the quantum correction to the masses up to order $\epsilon$. 
The methodology was explained before. The mass receives contributions
from quantum fluctuations of the Higgs field and of the photon. The
contribution for trivial topology has to be subtracted out from
the previous sum and added to the counterterm contribution to the energy.
In order to manipulate these individually divergent
quantities we make use of the zeta-function regularization method.
Leaving out, for the time being, the contribution of $\delta
\lambda_{n=0}$, the   procedure  can be summarized by  the formula:
\be
\EC=\lim_{s \longrightarrow 0} \left[ {\cal E}_\phi^{(0)}(s) + {\cal
E}_A^{(0)}(s) -{\cal E}_\varnothing^{(0)}(s)
+ \epsilon ({\cal E}_\phi^{(1)}(s) + {\cal E}_A^{(1)}(s) -{\cal
E}_\varnothing^{(1)})(s) \right] + \mathcal{O}(\epsilon^2)
\ee
If the right-hand side is analytic for $\Re(s)\ge 0$, then we can
say that the procedure is unambiguous and the quantum mass finite. 
This indeed turned out to be the case for the critical value of 
the area ($\epsilon=0$) as found in section~\ref{s.critical}. 
The corresponding limit is $\EC^{(0)}$ given in
Eq.~\ref{crit_energy}.
Our goal now is to see if this continues to be the case up to order
$\epsilon$.

The main formulas for obtaining the expansion of the analytically
continued energies ${\cal E}_Q(s)$ were explained earlier. In
particular, given the expansion of the eigenvalues of the fluctuation
operator $\lambda_i= \lambda_i^{(0)} + \epsilon \lambda_i^{(1)}+ \ldots$
one obtains the order epsilon correction to ${\cal E}_Q(s)$ 
as follows:
\be
\label{zetacorrec}
{\cal E}_Q^{(1)}(s)= -\frac{1}{2\Gamma(s-1/2)} \int_0^\infty dx\,
x^{s-1/2} \sum_i e^{-x \lambda_i^{(0)}} \lambda_i^{(1)}
\ee
However, in calculating the eigenvalues in Appendix~\ref{calculation}
we factored out from the operator a coefficient $\frac{e v}{\kappa}$. 
As explained earlier, $ev$ are just the natural units of quantum
energies, and it seems more reasonable  to analytically continue a dimensionless
expression and put the units back  at the end. Eliminating the factor $1/\kappa$ 
was dictated only by simplicity, since it is trivial to correct for it in
the final result:
\be
{\cal E}_Q(s) \longrightarrow \kappa^{s-1/2}{\cal E}_Q(s)=
(1+(s-1/2)\epsilon+\ldots){\cal E}_Q(s)
\ee
The modification  does not alter the cancellations between different
terms, but induces an order
$\epsilon$ correction proportional to the leading order result:
\be
\EC=(1-\frac{\epsilon}{2})\EC^{(0)} + \epsilon \bar{\cal
E}^{(1)} +\ldots
\ee
The new term $\bar{\cal E}^{(1)}/(ev)$ can be constructed using 
Eq.~\ref{zetacorrec} and the eigenvalues computed in
Appendix~\ref{calculation}
and given earlier in this section.  Its evaluation 
is the goal of the rest of this section. 

Let us start by computing the contribution of photon fluctuations
$\bar{\cal E}_A^{(1)}(s)$. 
Plugging the correction to the eigenvalues in the general formula one
gets:  
\be
\bar{\cal E}^{(1)}_A(s)/(ev)=-\frac{1}{2 \Gamma(s-1/2)} \int_0^\infty dx\,
x^{s-1/2} 
\, ({\cal F}(x/q)-1)=-\frac{q^{s+1/2}}{2 \Gamma(s-1/2)} \int_0^\infty dx\,
x^{s-1/2} \, ({\cal F}(x)-1)
\ee
The  integral is divergent at $s\le 1/2$. As done before, we   may regulate it
by adding and subtracting $e^{-x}/x$ to ${\cal F}$. With this
subtraction the integral converges down to  $s=0$. Curiously the
singularity at $s=1/2$ of the subtracted term is cancelled by the pole in the  Gamma
function, so that the whole expression is finite down to $s=0$:
\be
\bar{\cal E}^{(1)}_A(s)/(ev)=-\frac{q^{s+1/2}}{2}-\frac{q^{s+1/2}}{2 \Gamma(s-1/2)}
\int_0^\infty dx\,
x^{s-1/2} \, ({\cal F}(x)-1-\frac{e^{-x}}{x})
\ee

Now let us compute the contribution from the  Higgs sector $\bar{\cal E}_\phi^{(1)}$:
\be
\bar{\cal E}_\phi^{(1)}(s)/(ev)=-\frac{1}{\Gamma(s-1/2)} \int_0^\infty dx\,  x^{s-1/2} \,
\sum_{n=1}^{\infty}  e^{-xn} \big(qn-\sum_{j=0}^{n-2} \rho(j)\big)
\ee
We can treat the first term in the integral using: 
\be
\sum_{n=1}^\infty n e^{-xn} =-\frac{d}{dx} \sum_{n=1}^\infty
e^{-xn}=-\frac{d}{dx} \frac{1}{1-e^{-x}}=\frac{e^{-x}}{(1-e^{-x})^2} 
\ee
For the second term we use the explicit expression for $\rho(j)$ and 
exchange the sums in $j$ and $n$ as follows 
\be
e^{-\xi}\sum_{j=0}^\infty\sum_{n=j+2}^\infty \frac{\xi^j}{j!}e^{-xn}=
e^{-\xi}\sum_{j=0}^\infty \frac{\xi^j}{j!} \frac{e^{-x(j+2)}}{1-e^{-x}}=
\frac{e^{-2x}}{1-e^{-x}}e^{-\xi(1-e^{-x})}
\ee
Resumming  over $\vec{k}$ we get:
\be
\bar{\cal E}_\phi^{(1)}(s)/(ev)=\frac{1}{\Gamma(s-1/2)}\int_0^\infty dx\,
x^{s-1/2} \frac{e^{-x}}{1-e^{-x}} \,
(e^{-x}\mathcal{F}((1-e^{-x})/q)-\frac{q}{1-e^{-x}})
\ee
Again the integral diverges. It is convenient as usual to rearrange
the integrand subtracting  the leading behaviour of
$\mathcal{F}$ for small values of its argument. The remaining piece 
contains the divergent integral and takes the form:
\be
-\frac{q}{\Gamma(s-1/2)}\int_0^\infty dx\,
x^{s-1/2} \frac{e^{-x}}{1-e^{-x}} = -q
\frac{\Gamma(s+1/2)\zeta(s+1/2)}{\Gamma(s-1/2)} =-q (s-1/2) \zeta(s+1/2)
\ee
Notice that, as before, this term is regular since the factor $(s-1/2)$
cancels the pole of the Zeta function. 
We finally arrive to:
\be
\bar{\cal E}_\phi^{(1)}(s)/(ev)=-q (s-1/2) \zeta(s+1/2)+\frac{1}{\Gamma(s-1/2)}\int_0^\infty dx\,
x^{s-1/2} \frac{e^{-2x}}{1-e^{-x}} \,
(\mathcal{F}((1-e^{-x})/q)-\frac{q}{1-e^{-x}})
\ee
which is regular at all values of $s$. 

The next step is to subtract  from the previous terms 
the order $\epsilon$ contribution to  the vacuum energy 
for trivial topology. This is a simple matter since in
section~\ref{s.critical} we computed this energy for arbitrary values
of the area. From it, we can separate out the part which is
proportional to the critical area result and we are left with:
\be
\bar{\cal E}^{(1)}_\varnothing(s)/(ev)=-\frac{3q}{2} - \frac{3
q^{s+1/2}}{2
\Gamma(s-1/2)} \int dx \, x^{s-1/2} e^{-x q }
\left(\mathcal{F}(x)-\frac{1}{x}\right)
\ee
Once more the result is analytic for all values of $s$ (we recall that
Euler gamma function has no zeroes).

In summary, we have verified that the total contribution can be
analytically continued to the physical point $s=0$. Contrary to the
leading order case, each term is analytic by itself. We expect this to
happen at higher orders as well. Combining all factors, we arrive at
\bea
\nonumber
\bar{\cal E}^{(1)}/(ev) &=& \left(\frac{3
}{2}+\frac{\zeta(1/2)}{2}\right)q -\frac{\sqrt{q}}{2}+\\
\label{ebarone}
\frac{\sqrt{q}}{4 \sqrt{\pi}}&\int_0^\infty dx&\, \frac{1}{\sqrt{x}}
\left[(1-3e^{-xq})\mathcal{F}(x) -1
-\frac{(e^{-x}-3e^{-xq})}{x}\right]-\\
\nonumber
\frac{1}{2 \sqrt{\pi}}&\int_0^\infty dx&\,
 \frac{e^{-2x}}{\sqrt{x}(1-e^{-x})} \,
(\mathcal{F}((1-e^{-x})/q)-\frac{q}{1-e^{-x}})
\eea
This quantity has a dependence on $q$, which is explicitly displayed, and
a dependence on the metric (or the torus periods) hidden in the 
function $\mathcal{F}$. 

Now we are ready to present the final result up to order $\epsilon$ given 
by 
\be
\label{final}
{\cal E}= {\cal E}^{(0)} + \frac{\sqrt{\epsilon}}{2}+ \epsilon (\bar{\cal
E}^{(1)} -\frac{1}{2}\EC^{(0)}+\ER^{(1)})
\ee
The second term comes from the correction to the $n=0$ eigenvalue,
which was linear in $\epsilon$. Its contribution to the mass is  finite,
and can be added to the final result without any analytical continuation. 
The contribution $\ER^{(1)}$ depends on the scheme. In our scheme,
presented  in Appendix~\ref{s.renormalization}, the result is
$-\frac{13}{32}$. 

We may numerically evaluate the result  to explore its dependence on  $q$ and $\tau$.
For $q=1$ and $\tau=e^{i\pi/3}$ (triangular
lattice), at which the quantum energy had its
minimum at critical area, we get 
\be
{\cal E}/(ev)= 0.628039-\frac{39}{32} + 0.5 \sqrt{\epsilon}
-\epsilon\,  (0.1542505+\frac{13}{32})
\ee
where we have separated out the contributions of the Casimir energy
and the renormalization counterterm. 
On the other hand, for $\tau=i$ (square lattice) one gets
\be
{\cal E}/(ev)= 0.628873 -\frac{39}{32} + 0.5 \sqrt{\epsilon}
-\epsilon\, (0.155933+\frac{13}{32})
\ee
It is interesting to notice that the epsilon term goes in the
direction of compensating the angle dependence. This is what is
expected, since for large areas ($\epsilon \longrightarrow 1$) 
the dependence on the metric parameter $\tau$
should  disappear. With our numbers we see that for $\epsilon\sim 1/2$
we get a common value of the mass of $−0.5174$. We can take this number
as an crude estimate of the vortex mass on the plane. Alternatively, 
we might take the value obtained at $\epsilon=1$ which is $-0.652$.  

One can also explore the behaviour for large $q$. The leading
dependence goes linearly with $q$, as for the critical area result.
Dividing out the linear term by $q$, we get an energy per vortex 
equal to
\be
{\cal E}/(ev)=  (-\frac{7}{32}+\zeta(-1/2)+
\epsilon(\frac{19}{32}+\frac{-\zeta(-1/2)+\zeta(1/2)}{2})=
−0.42663- \epsilon\, 0.03249 
\ee
Although, it is not possible to draw any rigorous conclusion from 
our 2 terms of the expansion, we see that all estimates give a value
close to $-0.5 ev$ for the quantum mass of the vortex on the plane. 
For comparison with other numerical estimates, one  should guarantee
that the same renormalization prescription is adopted.

\section{Conclusions and Outlook}
\label{s.conclusions}
In this paper we have studied the one-loop quantum correction to the masses 
and energies of vortices  in the 2+1 dimensional Abelian Higgs Model
formulated on the torus. For a critical value of the area ${\cal A}_c$ 
the result can be computed analytically. Away from this value an expansion 
in powers of $\epsilon=({\cal A}-{\cal A}_c)/{\cal A}$ can be 
set up, of which we have computed the linear correction. In our
formulation, the theory is defined  on a unit  square torus with 
constant metric tensor. A general metric of this type  depends on a 
conformal  factor and a complex number $\tau$. The conformal factor is 
directly proportional to the area, and measures the departure from the critical 
area case. On the other hand the modular parameter $\tau$ can be mapped, 
after a change of variables to an euclidean metric, onto the periods of 
the torus, i.e. their  aspect ratio and relative angle. Hence, our quantum 
masses are indeed  functions of the complex parameter $\tau$ and, as
expected,  are invariant under transformations of the modular group. 
Indeed, our analytic results are expressed in terms of integrals of the
two-dimensional Riemann theta function, having this property. 
For the  critical area case we showed that the minimal energy is achieved 
for a torus which matches with a triangular lattice of vortices. It is
remarkable that the quantum corrections induce a breaking of the classical
degeneracy towards a configuration consistent with the standard vortex 
lattice in  type II superconductors. 

Our methodology allows also to study quantum energies for multivortex 
configurations. One of the advantages, compared to other numerical
methods,
is that we can fix from the start the position of the individual
vortices,
given by the zeroes of the Higgs field. At the classical level, the energy
depends only on the number of vortices $q$ and not on their positions. 
There is no known reason  why this independence should be preserved at the 
quantum level. Thus,  there could be attraction or repulsion of 
vortices induced by quantum corrections. Our result, however, shows that 
the degeneracy is preserved up to first order in $\epsilon$. There is no
apparent symmetry underlying this degeneracy, so that it could still be 
broken  at higher orders in $\epsilon$. This is an important  conceptual 
issue  which could be hard to settle in a purely numerical fashion. Our
result here is analytic but only valid for the first two terms in the
expansion. We hope this point could encourage other authors 
to extend our result to higher orders.

Another interesting piece of information  is the dependence of the quantum
energy on the number of vortices $q$. Notice, however, that the
critical area scales with the number of vortices. Hence, one should
actually talk about a critical value of the vortex density. If we
scale the area and the relative distances among vortices at the same
rate, one should expect that for large areas the quantum energy scales 
linearly with $q$. Dividing the multi-vortex energies by the number of
vortices one gets another estimate of the vortex mass. For small
values of the area, this linear dependence is modulated by  
corrections of order $\sqrt{q}$, $1/\sqrt{q}$ and subleading.

Since vortices are exponentially localized objects it is quite
plausible that the expansion  can be extrapolated to infinite area, 
obtaining the  one-loop quantum energies for vortices on the plane.
Indeed, once the area is a few times larger than the vortex size the
effect of the periodic boundary conditions should be exponentially 
suppressed, which would suggest a fast approach. 
In Ref.~\cite{AGARAMOSI} we investigated the shape of the Bogomolny 
solutions themselves using a  Bradlow parameter expansion up to order
$\epsilon^{51}$. The result could be numerically extrapolated to infinite area 
(vortices on the plane) and  compared successfully with other 
approaches. The expansion of the solution is only part of the program
in computing quantum energies,  and getting to high orders in all the
steps certainly demands more  efficient  and powerful automatization 
techniques. From our two terms in the expansion, we have played the game 
of extrapolating to infinite volume. The results are certainly not
crazy but, due to the limited information involved, a  serious
comparison with other results lacks any rigour. A more interesting
comparison  could be  to use  the methodology of Ref.~\cite{Izquierdo} for 
vortices on the torus, and compare the results with our exact results. 
These might give a new measure of the numerical errors involved in that 
method.

Finally, we should comment that the idea of a Bradlow parameter
expansion is quite general and extends to other types of vortices, 
abelian and non-abelian~\cite{ramos_thesis} and in non-commutative
space~\cite{lozano1}-\cite{lozano2}. Indeed, the idea itself  emerged
from a related  type of expansion that occurs for four-dimensional Yang-Mills
theories~\cite{instantons}. It is also applicable in principle to
non self-dual vortices and to vortices in 3+1 dimension, for which the
mass turns into the string tension~\cite{Baacke}.
It is  quite plausible that our technique
can be extended to the computation  quantum corrections in all those cases.

\appendix
\section{Corrections to the eigenvalues of the fluctuation operator}
\label{calculation}
In this appendix  we will collect the details of the calculation of the
eigenvalues of the quadratic quantum fluctuation operator $\VO$ to
order $\epsilon$. The methodology is explained in
section~\ref{s.expansion}.  The potential energy of fluctuations 
has two terms.  The first term is given by 
\be
\label{firstterm}
\frac{1}{2\kappa} \int dx\, |\tilde{D}\varphi -i C \overline{\delta A} \varphi -i C
\phC \delta a|^2 
\ee
The second term is 
\be
\label{secondterm}
\frac{1}{2\kappa} \int dx\, (\Im(\tilde{\partial}^*\delta a) +
\frac{C}{2}(\phC^* \varphi+ \phC \varphi^*))^2
\ee
where $C=e v/\VZ$. The operator for fluctuations $\VO$, whose
eigenvalues we have to calculate,  can be read from the potential 
divided by $ev/(2\kappa)$. Leaving the factor $1/\kappa$ out is 
dictated by simplicity, since it is trivial to correct for it in the
final result. We can see in the previous expressions, that 
the background fields $\overline{\delta A}$ and $\phC$
appear divided by $\VZ$, cancelling out the multiplicative $\VZ$
appearing in their expression. Thus, since the dependence on all the
constants is obvious, we can simplify the calculation of the spectrum 
by  choosing units $ev=\VZ=1$ (implying $C=1$).

Furthermore, as explained in section~\ref{s.expansion},  we will
benefit from the fact that we do not need to get involved into
technicalities associated with degenerate perturbation theory, since
all we need is to sum of the eigenvalues within each sector which is
degenerate at leading order.

\subsection{Calculation of $\delta \lambda_n$}
In this subsection we will explain the calculation of $\delta
\lambda_n$ according to the formula Eq.~\ref{lambdan}. There are two
terms: one coming from the diagonal Higgs-Higgs part $\VO_{11}^{(1)}$,
and the second coming from the mixed terms $\VO_{12}^{(1/2)}= \VO_{21}^{(1/2)}$.

Let us start with the contribution of  the diagonal part. This comes
from  two terms in the fluctuation potential:
\be
\int dx\, \left[ (\Re(\phC^* \varphi))^2 +2 \Im((\tilde{D}
\varphi)^* \overline{\delta A} \varphi)\right]
\ee
To calculate the correction to the leading order eigenvalue $n$,
we have to replace $\varphi=(a_{n s}+ i b_{n_s})\Psi_{n,s}$ for $n$
and $s$ fixed. Since we are interested in the trace we only need to
the coefficients that multiply $a_{n s}^2$ and $b_{n s}^2$,  and we 
can drop all mixed terms. After a simple calculation we arrive at 
\be
\sum_s \int dx\, \left[ |\phC|^2 |\Psi_{n,s}|^2 + 4  \Im((\tilde{D}
\Psi_{n,s})^* \overline{\delta A} \Psi_{n,s})\right]\equiv \epsilon (\tilde{K}_n
+ K_n)
\ee

In order to perform the integral of the first term we make use of the
Fourier expansion of the product $\Psi^*_{n, s} \phC$:
\be
V_{n,s}\equiv  \Psi^*_{n, s}(x) \phC(x)= \sqrt{\epsilon} \sum_{\vec{k}} \sum_{s'}
c_{s'} X^{s,s'}_{n,0}(\vec{k}) e^{-2\pi i \vec{k}\vec{x}}
\ee
Now we can plug the Fourier expansion of $V_{n,s}$ and that of its
complex conjugate and perform the integration over $x$. 
Hence, we get
\be
\tilde{K}_n= \sum_{\vec{k}} \sum_s |\sum_{s'}
c_{s'} X^{s,s'}_{n,0}(\vec{k}) |^2 = \sum_{\vec{k}} \frac{\xi^n}{n!}
e^{-\xi}\equiv \rho(n)
\ee
Notice that due to the unitarity of $U(\vec{k})$ the dependence on the 
moduli parameters $c_s$ has dropped completely. All that was left was
the norm of $c_s$ which is fixed by the Bogomolny equation to be equal
to 1. 

Now we proceed to calculate the integral of the second term, by
applying the $\tilde{D}$ operator and substituting the Fourier
expansion of $\overline{\delta A}$:
\bea
\nonumber
\epsilon\,K_n= 4 \sqrt{n} \sum_s  \int dx \, \Im\left(\overline{\delta A} \Psi^*_{n-1, s}
\Psi_{n , s}\right)=  \\ 4 \sqrt{n} \sum_s \sum_{\vec{k}} \Im\left( e^{i
\alpha(\vec{k})} X_{n-1\,  n}^{s s }(\vec{k})
\overline{G}_1(\vec{k})\right) = 
-2\epsilon \sum'_{\vec{k}} e^{-\xi}
L^{(1)}_{n-1}(\xi)  u(\vec{k}) \mathrm{Tr}(U(\vec{k}))
\eea
where we recall that $L^{(1)}_{n-1}$ is a generalized Laguerre
polynomial and $u(\vec{k})=c_t^* U_{t s}(-\vec{k}) c_s$.  
The trace of the unitary matrix $U(\vec{k})$ imposes that the 
Fourier component $k_i$ should be a multiple of $q$. This 
restriction eliminates the dependence on the moduli parameters
contained in $u(\vec{k})$. Altogether, the result becomes 
\be
K_n= -2 q \sum'_{\vec{k}}  e^{-q^2 \xi}
L^{(1)}_{n-1}(q^2 \xi) 
\ee
where the primed sum runs over over all integer vectors excluding $\vec{k}=0$ .

There is an alternative evaluation of $K_n$ which turns out to be more
useful. This follows by going back to the definition of $K_n$ as an
integral involving the operator $\tilde{D}$ and integrating by parts, 
passing the operator $\tilde{D}^\dagger$ to act on the other factors.
We leave the details to the reader and give here the relation that one
gets:
\be
K_n =   K_{n+1}- \frac{4}{\epsilon} \sum_s \int dx\,  \Im(\tilde{\partial}^*
\overline{\delta A})\, |\Psi_{n,s}(x)|^2
\ee
The last term can be evaluated by using the second Bogomolny equation. 
The final relation is then 
\be
K_n =   K_{n+1} + 2(\tilde{K}_n-q)
\ee
Using our previous result,  we conclude
\be
K_n= -2 \sum_{m=0}^{n-1} (\rho(m)-q) 
\ee
valid for $n\ge 1$.  The two expressions of $K_n$ look very 
different but they can be verified to give the same numerical values. 
By the definition it is clear that $K_{n=0}=0$.

Now we need to compute the contributions to $\lambda_n$ coming from the 
mixed terms $\VO_{12}=\VO_{21}$. Our first step is then precisely to
give the matrix elements of this operator in the basis of eigenstates 
of the lowest order potential. This can be read out from the
corresponding terms in the potential 
\be
\label{mixed}
\Im\left( \int dx \,  (\tilde{D} \varphi)^*\, \phC \delta
 a\right) + \int dx \, \Re(\varphi^*\phC) \Im(\tilde{\partial}^*
 \delta a) 
\ee
where $\varphi$ has to be treated as before. On the other hand the
fluctuation of the photon field $\delta a$ has two components, labelled 
$\delta a_T$ and $\delta a_L$, corresponding to the gauge independent
and gauge dependent part. Operationally the separation follows by 
computing $(\tilde{\partial}^* \delta a)$. For the transverse part it
is purely imaginary, while for the longitudinal one it is real. Obviously,
the second term in Eq.~\ref{mixed} involves only the transverse
fluctuations, while the first term involves both types. For the
transverse modes it is convenient to perform an integration 
by parts similar to that done  before for the diagonal term.  After 
combining it with the second term we arrive  at 
\be
\label{mixedterms}
\Im\left( \int dx \,  (\tilde{D} \varphi)^*\, \phC \delta
 a_L\right) +\Im\left( \int dx \,  \varphi^* (\tilde{D}^\dagger
 \phC) \delta   a_T\right) 
\ee
which separates neatly the longitudinal and transverse contributions. 

Now we are ready to obtain the matrix elements of the mixed terms. 
We replace $\varphi=(a_{n,s}+ i b_{n,s})\Psi_{n,s}(x)$ 
and $\delta a_\eta= r^\eta_{\vec{k},\sigma'}
\chi^\eta_{\vec{k},\sigma'}(x)$, where $\eta=L,T$ and the basis
vectors $\chi^\eta_{\vec{k},\sigma'}(x)$ are given in
Eq.~\ref{chidef}. We recall that $\vec{k}$ is restricted to positive
values. Later on we will consider the contribution of $\vec{k}=0$. 
We will use the index $\sigma=+$ for the terms
proportional to $a_{n,s}$ and  $\sigma=-$ for the terms
proportional to $b_{n,s}$. With this notation we will write
\be
\langle n,s,\sigma| V^{(1/2)}_{12} |\vec{k}, \eta, \sigma'\rangle\equiv
A_{\eta}(n,s,\sigma;\vec{k},\sigma')
\ee
where $\eta$ should be replaced by $T$ and $L$ for the longitudinal
and transverse modes respectively.

For the calculation of the transverse modes we will also need
 $X_{n 1}$ given by Eq.~\ref{matrixel}, which evaluates explicitly to 
\be
X_{n 1}^{s s'}(\vec{k})= U_{s s'}(\vec{k}) i^{(n-1)} e^{i
(n-1)\alpha(\vec{k})} Y_{n 1}(\xi)=  U_{s s'}(\vec{k}) i^{(n-1)} e^{i
(n-1)\alpha(\vec{k})} \frac{\xi^{(n-1)/2}}{\sqrt{n!}} e^{-\xi/2} (n-\xi)
\ee
valid for any $n$, and which defines the real quantity $Y_{n 1}(\xi)$. 
With the previous definitions and substitution into the second term of 
Eq.~\ref{mixedterms} we get:
\be 
\label{mat_el_gi}
A_{T}(n,s,\sigma;\vec{k},\sigma')=
\frac{\sqrt{\epsilon} Y_{n 1}(\xi)}{\sqrt{2}}
\Im\left[ i^{n-1}
e^{-i\pi(1-\sigma)/4}e^{i\pi(1-\sigma')/4} 
(e^{in\alpha(\vec{k})} U^{s s'}(\vec{k})+\sigma'e^{in\alpha(-\vec{k})}
U^{s s'}(-\vec{k}))c_{s'}\right]
\ee

A similar calculation can be done for the longitudinal terms, given by
the first term of Eq.~\ref{mixedterms}.  The result is
\be
\label{mat_el_gd}
A_{L}(n,s,\sigma;\vec{k},\sigma')=  \frac{\sqrt{n \epsilon} Y_{n-1\,
0}(\xi)}{\sqrt{2}}
\Re\left[ i^{n-1}
e^{-i\pi(1-\sigma)/4}e^{i\pi(1-\sigma')/4}
(e^{in\alpha(\vec{k})} U^{s s'}(\vec{k})+\sigma'e^{in\alpha(-\vec{k})}
U^{s s'}(-\vec{k}))c_{s'}\right]
\ee
where $Y_{n\,0}(\xi)= \xi^{n/2} e^{-\xi/2}/\sqrt{n!}$.
The previous results are complicated and depend on the  moduli
parameters $c_s$, related to  the location of  the vortices (zeroes of the
background Higgs field). As we will see in a minute this dependence 
drops when computing the contributions to the vortex mass.

With the mixed matrix elements given before it is relatively
straightforward to compute the contribution to the Higgs field
eigenvalue of the form
$\VO_{12}^{(1/2)}(n-\VO^{(0)}_{2 2})^{-1}\VO_{21}^{(1/2)}$. 
The contribution coming from  transverse photons becomes
\be
\label{transphot}
\sum_{\vec{k}>0} \sum_{s \sigma \sigma'}
\frac{(A_{T}(n,s,\sigma;\vec{k},\sigma'))^2}{(n-\xi)}=
\epsilon \sum_{\vec{k}>0} \frac{ 2 Y^2_{n\, 1}(\xi)}{n-\xi}=
\epsilon(\rho'(n-1)-\rho'(n))
\ee
where  $\rho'(n)$ is the same value as $\rho(n)$
excluding the $\vec{k}=0$ contribution:
\be
\rho'(n) =\rho(n) -\delta_{n 0}
\ee
The formula~\ref{transphot} is valid for $n=0$ if we take $\rho'(-1)=0$.
As mentioned previously, the computation  simplifies considerably due to 
summation over  $s$ and $\sigma$, which is all we need for the vortex
mass calculation.  In particular,  $\sigma$ appears as a factor
$e^{-i\pi(1-\sigma)/4}$ multiplying the remaining argument of the
imaginary part (which we call $W$ for simplicity). The sum then  operates 
in Eq.~\ref{transphot} as follows:
\be
 (\Im(W))^2+(\Im(i W))^2= |W|^2   
\ee
Hence, all phases drop from $W$. The sum over $\sigma'$ then 
simplifies as follows
\be
\sum_{\sigma'} |W(\vec{k})+\sigma'W(-\vec{k})|^2= 2 (|W(\vec{k})|^2+
|W(-\vec{k})|^2) 
\ee 
cancelling  the phase proportional to $\alpha(\vec{k})$. Finally, the sum
over $s$, the unitarity of $U_{s s'}(\vec{k})$ and the normalization 
of $c_s$ removes all dependence on the vortex locations and produces 
a fairly simple result. 

Repeating the same steps we can obtain the contribution of the longitudinal 
modes, given by 
\be
\sum_{\vec{k}>0} \sum_{s \sigma \sigma'}
\frac{(A_{L}(n,s,\sigma;\vec{k},\sigma'))^2}{
n}= \epsilon \sum_{\vec{k}>0} \frac{ 2n Y^2_{n-1\, 0}(\xi)}{n} =
\epsilon \rho'(n-1)
\ee
valid for all values of $n$ with the prescription   $\rho'(-1)=0$
adopted earlier.  

To complete the calculation of the mixed terms we need to consider the
contribution  of  the $\vec{k}=0$ states.  It comes from an expression 
similar to that involving longitudinal photons and is proportional 
to $X_{n-1\, 0}(\vec{k}=0)$, which vanishes for all values except $n=1$.
After a little bit of algebra one obtains that the 
contribution is  $2 \epsilon \delta_{n 1}$.

Combining everything together we get the final result for the Higgs
correction
\be
\delta \lambda_n = \epsilon\,(2 q n -2 \sum_{j=0}^{n-2} \rho(j)
+\delta_{n 0})
\ee
The second term vanishes for $n<2$. Notice that the $q$ dependence is
explicit in the first term, but is also present in the second term
through the metric. Finally, we stress that there is, indeed, a
correction to the eigenvalues at $n=0$. One can look into the details
of the calculation to see that there is only one of the $2q$
eigenvalues, which acquires a correction. The remaining zero-modes can
be obtained by differentiating the Bogomolny solution with respect to $c_s$. 
All variations orthogonal to $c_s$ should correspond to a zero-mode
since the energy does not depend on $c_s$. Variations of the type
$\varphi=i\phC$   decouple entirely from the potential. This is
so, because they correspond to global gauge transformations
$\phi \longrightarrow e^{i\theta}  \phi= \phi +i\theta \phi+\ldots$. 
We are only left with variations of the type $\phi=a_0\phC$ with 
$a_0$ real. Since the vector $c_s$ is normalized to 1, a re-scaling
does not correspond to a new solution. Hence, we do not have an
associated zero-mode, and its  contribution is captured by the trace and 
is equal to $\epsilon$ at  this order.

\subsection*{Calculation of $\delta \lambda_{\vec{k}}$}
We are now ready to calculate the corrections to the photon
eigenvalues. For transverse photons of momentum $\vec{k}$, the 
leading order eigenvalue was $\xi$. We will now evaluate the correction 
$\delta \lambda_{\vec{k}}$. As for the Higgs field there is a contribution
coming from diagonal parts $\VO_{2 2}^{(1)}$. This can be read out
from the potential of fluctuations
\be
\int dx \, |\phC|^2 |\delta a|^2
\ee
Now we should replace $\delta a= \chi^T_{\vec{k},\sigma'}$ 
for $\vec{k}>0$. Summing over the two values of $\sigma'$ and 
using the normalization of the background field $\phC$, the
result is just equal  to $2\epsilon$. 

The off-diagonal contribution can be easily evaluated using the matrix 
elements computed earlier. The result vanishes:
\be
\sum_{n} \sum_{s \sigma \sigma'}
\frac{(A_{T}(n,s,\sigma;\vec{k},\sigma'))^2}{ (\xi-n)}=
\epsilon \sum_{n} \frac{ 2 Y^2_{n\, 1}(\xi)}{\xi-n}=0
\ee
so that the final result is 
\be
\delta \lambda_{\vec{k}}= 2 \epsilon
\ee

We might also investigate the correction to the eigenvalue
corresponding to longitudinal photons. To leading order the 
eigenvalue vanishes since the potential is gauge invariant 
and the longitudinal photons are just gauge transformations. 
To order $\epsilon$ the contribution coming from the diagonal 
term is equal to $2\epsilon$, as for transverse photons. The off-diagonal 
contribution can be easily evaluated, giving 
\be
\sum_{n} \sum_{s \sigma \sigma'}
\frac{(A_{L}(n,s,\sigma;\vec{k},\sigma'))^2}{ (-n)}=
 \epsilon \sum_{n} \frac{ 2 n Y^2_{n-1\, 0}(\xi)}{-n}=-2\epsilon 
\ee
Hence, the sum of both  contributions vanishes as it should, since 
the argument of gauge invariance is valid for all values of $\epsilon$.

A similar conclusion can be drawn for the corrections to the zero
momentum photon field $\vec{k}=0$.  If we evaluate the correction to
these zero-modes, there are two contributions which cancel each other. 
Again this result follows from a symmetry that is valid for all values 
of $\epsilon$:  translation invariance (which is part of the moduli of 
classical solutions).

\section{Renormalization of the lagrangian parameters}
\label{s.renormalization}
In this appendix we study the remormalization of the lagrangian
parameters and its contribution to the vortex mass. The main point is 
that, in order for the physical quantities to have finite values, we
need to consider that the parameters in the lagrangian are not equal 
to the physically defined  quantities. Thus, we are led to consider 
the initial lagrangian with the constants $e,v,\lambda$ replaced by 
their bare  values $e_B , v_B, \lambda_B$. In general, if one wants to
construct   finite Green  functions one also needs a multiplicative
renormalization of the fields:
$\phi \longrightarrow \phi^B=Z_{\phi} \phi$ and $A_\mu
\longrightarrow A^B_\mu= Z_A A_\mu$. 
Within a given regularization method,  the physical observables 
expressed in terms of the bare parameters seem  divergent as the
regularization is removed. However, when replacing the bare parameters 
in terms of the renormalized ones the divergence cancels out and one 
gets a finite result. Thus, in order to implement the procedure one needs 
to express the bare quantities as a function of  the renormalized ones. 
Typically, what one does is to select a set of  physical quantities
and impose   that they  take a prescribed
form (usually the result of lowest order perturbation theory) as a function 
of the physical parameters. If the number of selected  quantities is
equal to the number of  bare constants,   one can solve for the latter in
terms of the physical (renormalized) ones and the cut-off. These equations
define the renormalization prescription or renormalization scheme. 
The final expression of any physical quantity  as a function of  the renormalized
constants is different for each prescription, but the difference can be 
accounted for by regarding the numerical value of the constants as scheme dependent. 
Very often the adopted prescription is dictated by simplicity (minimal 
substraction scheme, etc) at the expense of making the relation with 
physical quantities more complex. This is the renormalization idea in a 
nutshell. Now let us apply these ideas to our particular case.

From the previous perspective, the manipulations done in
section~\ref{s.method}, in which we did not take  renormalization into 
account, are valid  by simply replacing the coupling constants by their
bare counterparts. There are just  two places in which 
we assumed particular values of the constants: $\lambda=1$ and  
$\sqrt{|g^{(0)}|}=\frac{4 \pi q}{e^2 v^2}$. It makes no sense 
to replace these identities by the corresponding ones in terms 
of bare couplings. Thus, we have to maintain these relations as valid 
for the renormalised constants.

The theory is superrenormalizable in 2+1 dimensions so that the set of
divergent diagrams is very small. Thus, it is possible to impose  
simple renormalization  conditions on non-divergent quantities. 
In particular, it could be possible to assume no renormalization of 
the fields $Z_\phi=Z_A=1$. Nevertheless,  if we focus upon  physical
quantities as the energies, the results are not affected by the
renormalization of the fields.  Thus, we will consider here only the
renormalization of the constants of the theory: $\lambda=1$, $v$ and
$e$. Hence, we should replace them by the corresponding bare
quantities in the initial lagrangian: $\lambda_B=1+\delta\lambda$,
$v_B^2=v^2+\delta v^2$ and $e_B=e+\delta e$. Substituting these
expressions into the bare potential  density given in
Eq.~\ref{integrand}, we recover the original
potential  involving the renormalized constants, plus a correction 
linear  in $\hbar$, which is termed the counter-term potential:
\be
 \label{counter}
 -\sqrt{ |g|}\delta e \Im(A^*D\phi) +\delta e \frac{B |\phi|^2}{2}
-\delta v^2 \frac{e^2 \sqrt{ |g|}}{4} (|\phi|^2-v^2)+  \delta(\lambda
e^2) \frac{ \sqrt{|g|} }{8}
  (|\phi|^2-v^2)^2 
\ee
The expression is valid for any value of the flux and of the metric.
Notice that we have combined the renormalization of $\lambda$ and $e$
into the combination $(\lambda e^2)$ which appears more natural.

Starting from this point one can repeat the procedure to find the
quantum Casimir energy of the vortices: find the fields that minimize 
the classical energy and expand around them up to quadratic
fluctuations. Since the only modification is the addition of the
counterterm, which is of order $\hbar$, there is no modification in
the calculation of the quantum Casimir energies. However, there is an 
additional contribution to the  energy coming from integral of
the counterterm Eq.~\ref{counter} evaluated at the classical values 
of the fields. In our case, for $q=0$ these classical fields are $B=0$
and $\phi=v$, so that the correction vanishes. For  $q\ne 0$, using
the Bogomolny equations, the result is extremely simple:
\be
\frac{\delta(e v^2)}{2} \int dx\, B + \frac{\delta
\lambda}{2 \mathcal{A}} \int dx\, B^2
\ee
To evaluate the integrals  we use our parameterization  $B=\frac{2 \pi
q}{e}+ \delta B$, where $\delta B$ is of order $\epsilon$ and 
$\int dx\,  \delta B=0$. The resulting counter-term contribution 
up to order $\epsilon$ becomes
\be
\label{erenF}
\ER= \pi q \left( \frac{\delta(e v^2)}{e} + \frac{v^2 \delta
\lambda}{2}(1-\epsilon)\right)= \pi q \left( \delta v^2+ \frac{v^2
\delta(\lambda e^2)}{2 e^2}(1-\epsilon)
+\epsilon\frac{v^2\delta e}{e}\right)
\ee
valid in all schemes. Notice that we have expressed the result in terms 
of the coefficients of the different terms in the counterterm
potential. The choice of these coefficients depends on the
renormalization scheme. In principle, only $\delta v^2$ gets
contributions from divergent diagrams, so one could take $\delta
e=\delta \lambda=0$. Furthermore, given that the renormalization of
the parameters has to do with the ultraviolet properties of the 
theory, one could select a renormalization scheme in which the
coefficients are independent of the boundary conditions (hence on $q$)
and of  the metric tensor. With this choice, the counterterm contribution 
to the mass is linear in $q$ and 
independent of the metric. In the following paragraphs we will present 
a particular renormalization scheme based on the effective potential, 
which employs much of the machinery used earlier to compute the Casimir 
energy.

Our prescription is based on the computation of the dependence of the 
energy on external background fields. If we choose space-time
independent fields, what we are actually computing is the effective 
potential of the theory. On the basis of the previous considerations 
we will perform the calculation in the sector with trivial topology 
($q=0$) and for large values of the area. Let us begin by taking a
vanishing background vector potential and a real and constant Higgs
field $\phi(x)=\chi$. Although this seems to be gauge dependent, the 
final result  is actually the same if we transform the background field by an
arbitrary gauge transformation. 

The effective potential, just as the energy, is the sum of three
contributions. First of all, we have the classical potential. which 
for our background field becomes
\be
V_{(0)}(\chi)=  \frac{e^2\mathcal{A}}{8}(\chi^2-v^2)^2
\ee
The remaining two contributions start at  order $\hbar$. One is 
precisely the counterterm potential Eq.~\ref{counter} evaluated at the 
background field, whose form is: 
\be
\delta V(\chi)= 
-\delta v^2 \frac{e^2 \mathcal{A}}{4} (\chi^2-v^2)+
\delta(\lambda e^2)
  \frac{\mathcal{A}}{8}
    (\chi^2-v^2)^2    
\ee
The final contribution is obtained by integrating out quadratic quantum
fluctuations around the background field. No gauge fixing is performed 
on the fluctuation fields, explaining why the result is gauge invariant. 
Analyzing the quadratic  form, one sees that  the photon acquires a mass 
equal to $e \chi$,  while the Higgs field gets a mass
square equal to $\frac{3e^2}{2}\chi^2-\frac{e^2}{2}v^2\equiv e^2 v^2
\mu^2 $. With these considerations it is trivial to compute the
resulting quantum potential contribution
\be
V_Q(\chi)= \sum_{k_1=-\infty}^\infty
\sum_{k_2=-\infty}^\infty
\sqrt{4 \pi^2||\vec{k}||^2 +e^2 \chi^2}
+\frac{1}{2}\sum_{k_1=-\infty}^\infty
\sum_{k_2=-\infty}^\infty
\sqrt{4 \pi^2||\vec{k}||^2 +e^2 v^2\mu^2}
\ee
The quantity can be treated, as for the quantum energy, by analytical
continuation in the complex variable $s$. We essentially repeat the
steps that we employed in Eq.~\ref{calc_vac}. For example, for the
Higgs field fluctuation part we get 
\be
\frac{e v}{2 \Gamma(s-1/2)}\int_0^\infty x^{s-3/2} e^{-x\mu^2}
\mathcal{F}(\frac{4 \pi x}{e^2 v^2 \mathcal{A}}) 
\ee
Neglecting terms which are exponentially suppressed for large values
of the area,
the result becomes
\be
\frac{e^3 v^3 \mathcal{A}}{8 \pi}\frac{(\mu^2)^{3/2-s}}{s-3/2}
\ee
Adding the contribution of the photon fluctuations obtained along
similar lines we obtain 
\be
V_Q(\chi,s)=\frac{e^3 v^3 \mathcal{A}}{8 \pi(s-3/2)} (\mu^2)^{3/2-s} + 2
(\frac{\chi^2}{v^2})^{3/2-s}
\ee
If we set $\chi=v$, we recover the full calculation of the ground state quantum
energy in the trivial topology sector. On the other hand, we 
may  set $y=\frac{\chi^2}{v^2}-1$  and recall that $\mu^2=1+3y/2$ to
obtain a   series expansion of the previous formula in powers of $y$: 
\be 
\frac{e^3 v^3 \mathcal{A}}{8 \pi(s-3/2)} (3 +(s-3/2)(-\frac{7}{2}y +
\frac{17}{8}(s-1/2)y^2+ \ldots)
\ee
Notice that all but the first term are analytic at any value of $s$. 
We may now set $s=0$ and add it to the classical potential and
counterterm potential, which are both polynomials of degree 2 in $y$. 
Our renormalization conditions amount to imposing that the coefficients 
in $y$ and $y^2$ are given by the naive  potential. Hence, the contribution 
of the counterterm must exactly cancel the effective potential coefficients.
This gives two equations  which allow us to fix two of the
renormalization parameters.
\be
\label{renorm_param}
\delta v^2=-\frac{7}{4 \pi }e v \ ; \quad \quad \delta (\lambda e^2)=
\frac{17}{16 \pi}\frac{e^3}{v}
\ee

To fix the remaining renormalization of the charge, we consider a
different background field configuration. This time we take $\phi(x)=v$
and $A_i(x)=\tilde{V}_i$. Repeating the same procedure as before, we
compute the effective potential to be 
\be
\frac{v^2\mathcal{A}}{2}|\tilde{V}|^2 (e^2 +\delta(e^2))+ V_Q(\tilde{V})
\ee
We skip the details of the calculation of the quantum energy
$V_Q(\tilde{V})$ which is elaborate but straightforward. The 
coefficient of $|\tilde{V}|^2$ turns out to be 
\be
-\frac{e}{4 v} \sum_{\vec{k}} (||\vec{p}||^2 +1)^{3/2} \longrightarrow
-\frac{e^3 v \mathcal{A}}{8 \pi}
\ee
with $\vec{p}=2 \pi \vec{k}/(ev)$. The sum is convergent, and in the
large area limit tends to the expression on the right. Now imposing  the
renormalization condition that the coefficient of  $|\tilde{V}|^2$ 
equals the classical result, we find 
\be
\delta e= \frac{e^2}{8 \pi v}
\ee
which completes our renormalization  of parameters. 
The last step is to substitute these results onto the counterterm 
contribution to the quantum energy:
\be
\label{ERFinal}
\ER = q ev\left(-\frac{7}{4}+ \frac{17}{32}(1-\epsilon) +
\frac{\epsilon}{8}\right)= q
ev\left(-\frac{39}{32}-\epsilon\frac{13}{32}\right)\equiv
\ER^{(0)}+\epsilon \ER^{(1)}
\ee

\section*{Acknowledgments}
We acknowledge financial support from the MCINN grants
FPA2009-08785,
FPA2009-09017, FPA2012-31686, and FPA2012-31880, the Comunidad
Aut\'onoma de Madrid under the program  HEPHACOS
S2009/ESP-1473, and the European Union under Grant Agreement number
PITN-GA-2009-238353 (ITN STRONGnet). A.G-A participates in the project
Consolider-Ingenio 2010 CPAN (CSD2007-00042). We also benefit from the 
Centro de excelencia Severo Ochoa Program under  grant SEV-2012-0249.
Y.F. acknowledges financial support from Spanish MECD Grant
FIS2011-23713. 
\bibliography{references}

\begin{thebibliography}{25}
\expandafter\ifx\csname natexlab\endcsname\relax\def\natexlab#1{#1}\fi
\expandafter\ifx\csname bibnamefont\endcsname\relax
  \def\bibnamefont#1{#1}\fi
\expandafter\ifx\csname bibfnamefont\endcsname\relax
  \def\bibfnamefont#1{#1}\fi
\expandafter\ifx\csname citenamefont\endcsname\relax
  \def\citenamefont#1{#1}\fi
\expandafter\ifx\csname url\endcsname\relax
  \def\url#1{\texttt{#1}}\fi
\expandafter\ifx\csname urlprefix\endcsname\relax\def\urlprefix{URL }\fi
\providecommand{\bibinfo}[2]{#2}
\providecommand{\eprint}[2][]{\url{#2}}

\bibitem[{\citenamefont{Abrikosov}(1957)}]{abrikosov}
\bibinfo{author}{\bibfnamefont{A.~A.} \bibnamefont{Abrikosov}},
  \bibinfo{journal}{Sov Phys JETP} \textbf{\bibinfo{volume}{5}},
  \bibinfo{pages}{1174} (\bibinfo{year}{1957}),
  \urlprefix\url{http://www.mendeley.com/research/jetp51957-1174pdf/}.

\bibitem[{\citenamefont{Nielsen and Olesen}(1973)}]{Nielsen-Olesen}
\bibinfo{author}{\bibfnamefont{H.}~\bibnamefont{Nielsen}} \bibnamefont{and}
  \bibinfo{author}{\bibfnamefont{P.}~\bibnamefont{Olesen}},
  \bibinfo{journal}{Nuclear Physics B} \textbf{\bibinfo{volume}{61}},
  \bibinfo{pages}{45 } (\bibinfo{year}{1973}), ISSN \bibinfo{issn}{0550-3213},
  \urlprefix\url{http://www.sciencedirect.com/science/article/pii/0550321373903507}.

\bibitem[{\citenamefont{Taubes}(1980{\natexlab{a}})}]{taubes1}
\bibinfo{author}{\bibfnamefont{C.~H.} \bibnamefont{Taubes}},
  \bibinfo{journal}{Commun.Math.Phys.} \textbf{\bibinfo{volume}{75}},
  \bibinfo{pages}{207} (\bibinfo{year}{1980}{\natexlab{a}}).

\bibitem[{\citenamefont{Taubes}(1980{\natexlab{b}})}]{taubes2}
\bibinfo{author}{\bibfnamefont{C.~H.} \bibnamefont{Taubes}},
  \bibinfo{journal}{Commun.Math.Phys.} \textbf{\bibinfo{volume}{72}},
  \bibinfo{pages}{277} (\bibinfo{year}{1980}{\natexlab{b}}).

\bibitem[{\citenamefont{Jaffe and Taubes}(1980)}]{jaffe}
\bibinfo{author}{\bibfnamefont{A.}~\bibnamefont{Jaffe}} \bibnamefont{and}
  \bibinfo{author}{\bibfnamefont{C.}~\bibnamefont{Taubes}},
  \emph{\bibinfo{title}{Vortices and monopoles: structure of static gauge
  theories}}, Progress in physics (\bibinfo{publisher}{Birkh{\"a}user},
  \bibinfo{year}{1980}), ISBN \bibinfo{isbn}{9783764330255},
  \urlprefix\url{http://books.google.es/books?id=I2uOM8siLQ8C}.

\bibitem[{\citenamefont{Bogomolny}(1976)}]{Bogomolny}
\bibinfo{author}{\bibfnamefont{E.}~\bibnamefont{Bogomolny}},
  \bibinfo{journal}{Sov.J.Nucl.Phys.} \textbf{\bibinfo{volume}{24}},
  \bibinfo{pages}{449} (\bibinfo{year}{1976}).

\bibitem[{\citenamefont{de~Vega and Schaposnik}(1976)}]{devega}
\bibinfo{author}{\bibfnamefont{H.~J.} \bibnamefont{de~Vega}} \bibnamefont{and}
  \bibinfo{author}{\bibfnamefont{F.~A.} \bibnamefont{Schaposnik}},
  \bibinfo{journal}{Phys. Rev. D} \textbf{\bibinfo{volume}{14}},
  \bibinfo{pages}{1100} (\bibinfo{year}{1976}).

\bibitem[{\citenamefont{Weinberg}(1979)}]{weinberg}
\bibinfo{author}{\bibfnamefont{E.~J.} \bibnamefont{Weinberg}},
  \bibinfo{journal}{Phys. Rev. D} \textbf{\bibinfo{volume}{19}},
  \bibinfo{pages}{3008} (\bibinfo{year}{1979}).

\bibitem[{\citenamefont{Jacobs and Rebbi}(1979)}]{rebbi}
\bibinfo{author}{\bibfnamefont{L.}~\bibnamefont{Jacobs}} \bibnamefont{and}
  \bibinfo{author}{\bibfnamefont{C.}~\bibnamefont{Rebbi}},
  \bibinfo{journal}{Phys. Rev. B} \textbf{\bibinfo{volume}{19}},
  \bibinfo{pages}{4486} (\bibinfo{year}{1979}).

\bibitem[{\citenamefont{Gonzalez-Arroyo and Ramos}(2004)}]{AGARAMOSI}
\bibinfo{author}{\bibfnamefont{A.}~\bibnamefont{Gonzalez-Arroyo}}
  \bibnamefont{and} \bibinfo{author}{\bibfnamefont{A.}~\bibnamefont{Ramos}},
  \bibinfo{journal}{JHEP} \textbf{\bibinfo{volume}{0407}}, \bibinfo{pages}{008}
  (\bibinfo{year}{2004}), \eprint{hep-th/0404022}.

\bibitem[{\citenamefont{Manton}(1982)}]{manton}
\bibinfo{author}{\bibfnamefont{N.}~\bibnamefont{Manton}},
  \bibinfo{journal}{Physics Letters B} \textbf{\bibinfo{volume}{110}},
  \bibinfo{pages}{54 } (\bibinfo{year}{1982}), ISSN \bibinfo{issn}{0370-2693}.

\bibitem[{\citenamefont{Gonzalez-Arroyo and Ramos}(2007)}]{AGARAMOSII}
\bibinfo{author}{\bibfnamefont{A.}~\bibnamefont{Gonzalez-Arroyo}}
  \bibnamefont{and} \bibinfo{author}{\bibfnamefont{A.}~\bibnamefont{Ramos}},
  \bibinfo{journal}{JHEP} \textbf{\bibinfo{volume}{0701}}, \bibinfo{pages}{054}
  (\bibinfo{year}{2007}), \eprint{hep-th/0610294}.

\bibitem[{\citenamefont{Bordag et~al.}(2002)\citenamefont{Bordag, Goldhaber,
  van Nieuwenhuizen, and Vassilevich}}]{Bordag}
\bibinfo{author}{\bibfnamefont{M.}~\bibnamefont{Bordag}},
  \bibinfo{author}{\bibfnamefont{A.~S.} \bibnamefont{Goldhaber}},
  \bibinfo{author}{\bibfnamefont{P.}~\bibnamefont{van Nieuwenhuizen}},
  \bibnamefont{and}
  \bibinfo{author}{\bibfnamefont{D.}~\bibnamefont{Vassilevich}},
  \bibinfo{journal}{Phys. Rev. D} \textbf{\bibinfo{volume}{66}},
  \bibinfo{pages}{125014} (\bibinfo{year}{2002}),
  \urlprefix\url{http://link.aps.org/doi/10.1103/PhysRevD.66.125014}.

\bibitem[{\citenamefont{Vassilevich}(2003)}]{Vassilevich}
\bibinfo{author}{\bibfnamefont{D.~V.} \bibnamefont{Vassilevich}},
  \bibinfo{journal}{Phys. Rev. D} \textbf{\bibinfo{volume}{68}},
  \bibinfo{pages}{045005} (\bibinfo{year}{2003}),
  \urlprefix\url{http://link.aps.org/doi/10.1103/PhysRevD.68.045005}.

\bibitem[{\citenamefont{Izquierdo et~al.}(2005)\citenamefont{Izquierdo,
  Fuertes, Mayado, and Guilarte}}]{Izquierdo}
\bibinfo{author}{\bibfnamefont{A.~A.} \bibnamefont{Izquierdo}},
  \bibinfo{author}{\bibfnamefont{W.~G.} \bibnamefont{Fuertes}},
  \bibinfo{author}{\bibfnamefont{M.~d. l.~T.} \bibnamefont{Mayado}},
  \bibnamefont{and} \bibinfo{author}{\bibfnamefont{J.~M.}
  \bibnamefont{Guilarte}}, \bibinfo{journal}{Phys. Rev. D}
  \textbf{\bibinfo{volume}{71}}, \bibinfo{pages}{125010}
  (\bibinfo{year}{2005}),
  \urlprefix\url{http://link.aps.org/doi/10.1103/PhysRevD.71.125010}.

\bibitem[{\citenamefont{Izquierdo et~al.}(2004)\citenamefont{Izquierdo,
  Fuertes, Mayado, and Guilarte}}]{Izquierdo2}
\bibinfo{author}{\bibfnamefont{A.~A.} \bibnamefont{Izquierdo}},
  \bibinfo{author}{\bibfnamefont{W.~G.} \bibnamefont{Fuertes}},
  \bibinfo{author}{\bibfnamefont{M.~d. l.~T.} \bibnamefont{Mayado}},
  \bibnamefont{and} \bibinfo{author}{\bibfnamefont{J.~M.}
  \bibnamefont{Guilarte}}, \bibinfo{journal}{Phys. Rev. D}
  \textbf{\bibinfo{volume}{70}}, \bibinfo{pages}{061702}
  (\bibinfo{year}{2004}),
  \urlprefix\url{http://link.aps.org/doi/10.1103/PhysRevD.70.061702}.

\bibitem[{\citenamefont{Izquierdo et~al.}(2002)\citenamefont{Izquierdo,
  Fuertes, León, and Guilarte}}]{Izquierdo3}
\bibinfo{author}{\bibfnamefont{A.~A.} \bibnamefont{Izquierdo}},
  \bibinfo{author}{\bibfnamefont{W.~G.} \bibnamefont{Fuertes}},
  \bibinfo{author}{\bibfnamefont{M.~G.} \bibnamefont{León}}, \bibnamefont{and}
  \bibinfo{author}{\bibfnamefont{J.~M.} \bibnamefont{Guilarte}},
  \bibinfo{journal}{Nuclear Physics B} \textbf{\bibinfo{volume}{635}},
  \bibinfo{pages}{525 } (\bibinfo{year}{2002}), ISSN \bibinfo{issn}{0550-3213},
  \urlprefix\url{http://www.sciencedirect.com/science/article/pii/S0550321302003413}.

\bibitem[{\citenamefont{Baacke and Kevlishvili}(2008)}]{Baacke}
\bibinfo{author}{\bibfnamefont{J.}~\bibnamefont{Baacke}} \bibnamefont{and}
  \bibinfo{author}{\bibfnamefont{N.}~\bibnamefont{Kevlishvili}},
  \bibinfo{journal}{Phys. Rev. D} \textbf{\bibinfo{volume}{78}},
  \bibinfo{pages}{085008} (\bibinfo{year}{2008}),
  \urlprefix\url{http://link.aps.org/doi/10.1103/PhysRevD.78.085008}.

\bibitem[{\citenamefont{{Rebhan} et~al.}(2004)\citenamefont{{Rebhan}, {van
  Nieuwenhuizen}, and {Wimmer}}}]{Rebhan}
\bibinfo{author}{\bibfnamefont{A.}~\bibnamefont{{Rebhan}}},
  \bibinfo{author}{\bibfnamefont{P.}~\bibnamefont{{van Nieuwenhuizen}}},
  \bibnamefont{and} \bibinfo{author}{\bibfnamefont{R.}~\bibnamefont{{Wimmer}}},
  \bibinfo{journal}{Nuclear Physics B} \textbf{\bibinfo{volume}{679}},
  \bibinfo{pages}{382} (\bibinfo{year}{2004}), \eprint{arXiv:hep-th/0307282}.

\bibitem[{\citenamefont{Bradlow}(1990)}]{bradlow}
\bibinfo{author}{\bibfnamefont{S.}~\bibnamefont{Bradlow}},
  \bibinfo{journal}{Commun.Math.Phys.} \textbf{\bibinfo{volume}{135}},
  \bibinfo{pages}{1} (\bibinfo{year}{1990}).

\bibitem[{\citenamefont{Mumford}(1982)}]{tata}
\bibinfo{author}{\bibfnamefont{D.}~\bibnamefont{Mumford}},
  \emph{\bibinfo{title}{{ Tata Lectures on theta I}}}
  (\bibinfo{publisher}{{Birkhauser-Boston}}, \bibinfo{year}{1982}).

\bibitem[{\citenamefont{Ramos}(2007)}]{ramos_thesis}
\bibinfo{author}{\bibfnamefont{A.}~\bibnamefont{Ramos}},
  \emph{\bibinfo{title}{{ The Bradlow parameter expansion and its applications
  in field theory}}} (\bibinfo{publisher}{{PhD thesis, Univ. Autonoma de
  Madrid}}, \bibinfo{year}{2007}).

\bibitem[{\citenamefont{Lozano et~al.}(2006)\citenamefont{Lozano, Marques, and
  Schaposnik}}]{lozano1}
\bibinfo{author}{\bibfnamefont{G.}~\bibnamefont{Lozano}},
  \bibinfo{author}{\bibfnamefont{D.}~\bibnamefont{Marques}}, \bibnamefont{and}
  \bibinfo{author}{\bibfnamefont{F.}~\bibnamefont{Schaposnik}},
  \bibinfo{journal}{JHEP} \textbf{\bibinfo{volume}{0609}}, \bibinfo{pages}{044}
  (\bibinfo{year}{2006}), \eprint{hep-th/0606099}.

\bibitem[{\citenamefont{Lozano et~al.}(2007)\citenamefont{Lozano, Marques, and
  Schaposnik}}]{lozano2}
\bibinfo{author}{\bibfnamefont{G.}~\bibnamefont{Lozano}},
  \bibinfo{author}{\bibfnamefont{D.}~\bibnamefont{Marques}}, \bibnamefont{and}
  \bibinfo{author}{\bibfnamefont{F.}~\bibnamefont{Schaposnik}},
  \bibinfo{journal}{JHEP} \textbf{\bibinfo{volume}{0709}}, \bibinfo{pages}{095}
  (\bibinfo{year}{2007}), \eprint{0708.2386}.

\bibitem[{\citenamefont{Garcia~Perez et~al.}(2000)\citenamefont{Garcia~Perez,
  Gonzalez-Arroyo, and Pena}}]{instantons}
\bibinfo{author}{\bibfnamefont{M.}~\bibnamefont{Garcia~Perez}},
  \bibinfo{author}{\bibfnamefont{A.}~\bibnamefont{Gonzalez-Arroyo}},
  \bibnamefont{and} \bibinfo{author}{\bibfnamefont{C.}~\bibnamefont{Pena}},
  \bibinfo{journal}{JHEP} \textbf{\bibinfo{volume}{0009}}, \bibinfo{pages}{033}
  (\bibinfo{year}{2000}), \eprint{hep-th/0007113}.

\end{thebibliography}
\end{document}